\documentclass[conference]{IEEEtran}

\ifCLASSINFOpdf
\else
\fi
\usepackage{xcolor}
\usepackage{listings}
\usepackage{tcolorbox}
\tcbuselibrary{breakable, skins}
\usepackage{booktabs}
\usepackage{multirow}
\usepackage{graphicx} 
\usepackage{amssymb}  
\usepackage{pifont}   
\usepackage{xcolor}   
\usepackage{array}    
\usepackage{makecell}

\usepackage{tikz}
\usepackage{pgfplots}
\pgfplotsset{compat=1.18}
\definecolor{myblue}{RGB}{70,130,180}
\definecolor{myorange}{RGB}{255,165,0}
\definecolor{myred}{RGB}{220,20,60}
\definecolor{mypurple}{RGB}{128,0,128}
\definecolor{mygreen}{RGB}{60,179,113}
\usepackage{cite}       
\usepackage[hidelinks]{hyperref}

\newcommand{\cmark}{\textcolor{blue}{\checkmark}}
\newcommand{\xmark}{\textcolor{red}{\ding{55}}}

\newtcolorbox{promptbox}[1][]{
    enhanced,
    breakable,
    colback=black!5, 
    colframe=black!50, 
    title={\textbf{Prompt Template}}, 
    fonttitle=\bfseries,
    attach boxed title to top left={xshift=5mm, yshift=-2mm},
    boxed title style={
        colback=black!20,
        colframe=black!50,
    },
    #1
}

\newcommand{\toolname}{\textsc{FirmCure }}

\begin{document}
%
\title{\toolname: Towards Autonomous and Adaptive Rehosting of Linux-Based Firmware}

\author{
\IEEEauthorblockN{
Chuan Hong\textsuperscript{*}, Zheng Zhang, Lei Zhou\textsuperscript{*\textsection}, Laisong Li, \\
Chenyifan Liu, Ze Huang, Xu Zhou\textsuperscript{*\textsection}, and Peihong Lin
}
\IEEEauthorblockA{
\emph{National University of Defense Technology} \\
\{hongchuan24, zhangzhengnudt, zhoulcs, lilaisong, liuchenyifan, huangze, zhouxu, phlin22\}@nudt.edu.cn
}
}

	

%


\IEEEoverridecommandlockouts
\makeatletter\def\@IEEEpubidpullup{6.5\baselineskip}\makeatother
\IEEEpubid{\parbox{\columnwidth}{
        preprint
}
\hspace{\columnsep}\makebox[\columnwidth]{}}

\maketitle

\begin{abstract}
Full-system rehosting plays a critical role in the security analysis of Linux-based firmware. It matches commonly deployed firmware with sufficient background knowledge. However, for custom devices, existing approaches struggle to handle initialization and runtime obstacles in the rehosting process caused by specialized architectures and hardware-dependent configuration, which heavily rely on expert intervention. This ultimately creates fundamental bottlenecks and results in low rehosting efficiency.

To address the above challenges, we propose \toolname, the first LLM-driven full-system rehosting framework designed for autonomous and adaptive rehosting of Linux-based firmware. \toolname develops an Adaptive Perception Inference mechanism to extract firmware structural dependencies via static analysis, followed by a Reflective Synthesis module for iterative configuration optimization, and finally an Autonomous Runtime Intervention module for real-time error remediation through runtime fault diagnosis and monitoring. We evaluated 21 IoT firmware images from 10 vendors across 5 architectures, while \toolname achieved a 100\% network port opening rate and 90.5\% service interactivity, substantially outperforming state-of-the-art baselines. Our experiments confirm that \toolname's intervention strategies generalize across heterogeneous firmware. The framework successfully reproduces known vulnerabilities and discovers new security flaws.
\end{abstract}



%
\IEEEpeerreviewmaketitle

\section{Introduction}
The rapid development of Internet of Things (IoT), projected to exceed 50 billion devices by 2035\cite{iotanalytics2025connected}, necessitates urgent security analysis, particularly for Linux-based firmware comprising 58\% of devices\cite{commandlinux2024linux}. Firmware rehosting technology, as base for dynamic analysis platform, can keep the services running without physical hardware dependency \cite{zhou2023survey,Yin2021FirmHunter,zheng2019firm,srivastava2019firmfuzz}. Especially for full-system rehosting, it can maintain the kernel interactions and runtime context, making it applicable for system-level vulnerability analysis~\cite{angelakopoulos2024pandawan}. For example, Recent studies, such as FirmSolo~\cite{angelakopoulos2023firmsolo} and FirmDiff~\cite{angelakopoulos2024firmdiff} pursue kernel compatibility by modifying the kernel. In contrast, Firmadyne~\cite{chen2016firmadyne} and FirmAE~\cite{kim2020firmae} substitute the original kernel with a pre-built kernel while retaining the target filesystem and user-space applications. However, the above approaches adopt the controllable kernel deployment and fixed-ruled template mechanism to smoothly execute user-level applications, while avoiding complex system initialization issues caused by missing special hardware configuration. However, many issues remain unsolved, such as service crashes and deadlocks~\cite{qin2026firmwell}, which are still an unsolved challenge for firmware rehosting. In this work, we emphasize that the core obstacle of existing firmware rehosting approaches lies in case-specific and scenario-dependent complexities.

\textbf{First, complete initialization remains a key gap in firmware rehosting. }
The specialized hardware designs in custom devices are hard to be faithfully emulated, preventing existing approaches from achieving complete initialization. Full‑system rehosting approaches~\cite{chen2016firmadyne,kim2020firmae} treat the init routine as a black box and blindly execute, so any failure in a process unrelated to the target service can deadlock the entire initialization. User-mode rehosting approach Firmwell~\cite{qin2026firmwell} attempts to mitigate this by running the full init and reactively fixing failures, but its heuristic repairs cannot handle unanticipated faults such as complex hardware probing or blocking ioctls. As a result, neither approach can balance preserving essential service dependencies against avoiding hardware‑induced deadlocks, leaving the user‑space environment inherently fragile.

\textbf{Second, resolving rehosting issues in custom firmware remains heavily manual and fails to scale.}
The process depends on either time‑consuming manual intervention or pre‑built rules that frequently fail. Booting a new firmware image requires carefully matching QEMU\cite{bellard2005qemu} parameters to the target architecture and filesystem layout, yet a single wrong combination causes a crash whose unstructured log rarely points to the faulty setting. Existing Approaches cannot derive the correct configuration automatically from such opaque feedback. Even after a successful boot, services run into unpredictable errors like binary crashes, missing files, or endless polling loops. FirmAE~\cite{kim2020firmae} carries a fixed catalog of heuristic arbitration strategies, and Greenhouse~\cite{tay2023greenhouse} relies on predefined intervention measures, but both only cover failures their designers have already seen. When a crash or hang falls outside that known set, the tool simply halts and offers no further help. Analysts must then reverse engineer the firmware, manually try different parameters, and improvise workarounds, a cycle that is slow, labor‑intensive, and impossible to scale across diverse firmware images.

\textbf{Third, existing LLM-based approaches remain confined to static analysis and struggle with runtime rehosting issues.}
The emergence of large language models (LLMs)~\cite{radford2018improving,radford2019language,brown2020language} offers a new way to address the complexity of firmware rehosting, benefiting from their strong generalization ability to comprehend diverse hardware behaviors and runtime anomalies~\cite{shang2026empirical,ye2024detecting,pordanesh2024exploring}. Nevertheless, simply deploying raw LLMs cannot achieve reliable rehosting in practice. In real scenarios, precise fault localization and reasonable remediation strategy formulation rely on customized prompt design and dedicated toolchain orchestration. However, existing LLM-based approaches~\cite{lei2025flexemu} merely adopt LLMs for static knowledge extraction, such as parsing peripheral source code and modeling peripherals. They lack tailored customization and runtime tool cooperation, and thus fail to fully unleash the potential of LLMs in handling runtime failures.

To address these limitations, we propose \toolname, the first LLM-driven full-system rehosting framework for autonomous and adaptive rehosting of Linux-based firmware. \toolname  leverages the LLM-based agent’s ability to semantically understand heterogeneous initialization logic and its iterative reasoning cycle of planning, observation, and reflection~\cite{yao2022react,shinn2023reflexion,wei2022chain} to dynamically diagnose and repair runtime failures. Moreover, the agent draws on embedded Linux domain knowledge and invokes specialized tools for binary analysis, filesystem repair, and network configuration verification. 

\toolname organizes this intelligence into a closed-loop rehosting system structured around three sequential modules. The Adaptive Perception Inference module autonomously parses the firmware rootfs to extract architectural and service semantics, distinguishing hardware-dependent operations from critical service logic to synthesize a minimal context blueprint that isolates true execution dependencies. The Reflective Synthesis module translates this blueprint into a bootable environment through an iterative validation cycle, diagnosing boot failures from execution feedback and applying targeted repairs to kernel parameters or filesystem structures until kernel initialization and rootfs mounting are systematically stabilized. The Autonomous Runtime Intervention module continuously monitors post-boot execution states and dynamically routes runtime faults to specialized agents, resolving diverse anomalies through runtime binary execution bypass, environment remediation, and configuration adaptation to maximize progression across core rehosting metrics and continuously advance the rehosting state toward full service interactivity.

We evaluated \toolname on 21 firmware images from 10 vendors spanning 5 architectures. It achieves a 100\% network port activation rate and 90.5\% end-to-end service interactivity, substantially exceeding existing baselines. The multi-stage pipeline generalizes across heterogeneous targets, with each component independently resolving a distinct category of boot and runtime failures. \toolname reproduced all ten target CVEs and discovered five previously unknown vulnerabilities in commercial devices, confirming its utility for practical security assessment. 

In summary, we make the following contributions in this paper:
\begin{itemize}
    \item We propose \toolname, the first LLM-driven full-system rehosting framework that automates the entire rehosting lifecycle through a closed-loop pipeline of adaptive perception, reflective synthesis, and autonomous runtime intervention for Linux-based firmware.
    \item We design a multi-agent runtime intervention mechanism that autonomously monitors post-boot execution states, dynamically routes diverse runtime faults to specialized agents, and orchestrates remediation strategies that continuously advance rehosting metrics toward full service interactivity.
    \item We evaluate \toolname on 21 IoT firmware images across 10 vendors and 5 architectures, achieving a 100\% network port activation rate and 90.5\% service interactivity, substantially outperforming state-of-the-art baselines.
    \item We validate \toolname's practical security impact by successfully reproducing known CVEs and discovering unknown vulnerabilities in commercial devices, demonstrating its effectiveness for real-world firmware security analysis.
\end{itemize}

\section{Background} \label{sec:background}

\subsection{Firmware Rehosting} \label{sec:background_rehosting}

Firmware rehosting executes firmware images on a host via hardware emulation to enable dynamic analysis without physical devices~\cite{bellard2005qemu,chen2016firmadyne}. High-fidelity full-system rehosting of Linux-based firmware remains difficult because of heterogeneous hardware dependencies, driving state-of-the-art systems to prioritize restoring critical network services such as \texttt{httpd} and \texttt{sshd} for fuzzing~\cite{kim2020firmae,chen2016firmadyne,tay2023greenhouse,qin2026firmwell,zheng2019firm}. \textbf{User-mode Rehosting}, as adopted by Greenhouse~\cite{tay2023greenhouse} and Firmwell~\cite{qin2026firmwell}, translates a single process and delegates syscalls to the host kernel, bypassing the original kernel and network stack. \textbf{Full-system Rehosting}, as demonstrated by Firmadyne~\cite{chen2016firmadyne} and FirmAE~\cite{kim2020firmae}, virtualizes the entire platform, allowing the kernel to boot and network services to operate, making it the dominant choice for in-depth vulnerability analysis.

Within full-system rehosting, handling the kernel follows two strategies, \textit{Kernel Modification} and \textit{Kernel Replacement}. The modification strategy, used by FirmSolo~\cite{angelakopoulos2023firmsolo} and FirmDiff~\cite{angelakopoulos2024firmdiff}, reverse-engineers kernel configurations and corrects data-structure layouts, incurring heavy overhead from recompilation and manual alignment. The replacement strategy, pioneered by Firmadyne and extended by FirmAE, swaps the original kernel with a pre‑compiled, closely‑matching one while keeping user‑space intact. This avoids the overhead of kernel modification and enables scalable automated testing. Because perfect hardware fidelity is infeasible, rehosting follows a \textit{best-effort} principle, maximizing fidelity for security-critical components and gracefully degrading non‑essential dependencies, which is the core objective of \toolname.

\begin{table*}[t]
\centering
\caption{Failure Cases of Existing Firmware Rehosting Approaches and Their Manual Repair Strategies}
\label{tab:failure_cases}
\resizebox{\textwidth}{!}{%
\begin{tabular}{l l p{3.5cm} p{6cm} p{4.5cm} l}
\toprule
\textbf{Vendor} & \textbf{Device} & \textbf{Example} & \textbf{Failure Reason} & \textbf{Feasible Fix Strategy} & \textbf{Boot Stage} \\
\midrule
Totolink & NR1800X & Config Path Mismatch & \texttt{lighttpd} fails to locate config at non-standard path \texttt{/lighttpd/lighttpd.conf} & correct config path or create symlink & Service Runtime \\
\addlinespace
Tenda & AC18 & Service Early Exit & \texttt{ConnectCfm} routine invokes \texttt{cfmd} daemon, which terminates prematurely due to strict hardware dependencies & Bypass blocking calls via \texttt{gdb} breakpoints or dynamic instruction patching & Service Runtime \\
\addlinespace
TP-Link & TL-IPC43AN & Missing Runtime Flag & Infinite polling loop waiting for missing flag file \texttt{/tmp/jffs2\_ready} & Create file to break polling loop & Service Runtime \\
\addlinespace
TRENDnet & TEW-813DRU & Initialization Deadlock & \texttt{rcS} script deadlocks due to unresolved hardware configuration dependencies & Inject minimal synthetic \texttt{.cfg} to decouple checks & User-space Init. \\
\addlinespace
WAVLINK & NU516U1 & Emulation Parameter Error & Kernel panic caused by CPU architecture mismatch during QEMU launch & Adjust \texttt{-cpu 74Kf} in QEMU launch command & Kernel Bootstrap \\
\addlinespace
D-Link & DGL-5500 & Filesystem Corruption & Broken symbolic links cause rootfs mount failure & Remove dangling symlinks \& Restore missing libraries & Rootfs Mounting \\
\bottomrule
\end{tabular}%
}
\end{table*}

\subsection{Runtime Environment Reconstruction} \label{sec:background_workflow}

The kernel replacement approach shifts the core challenge to \textbf{runtime environment reconstruction}. Since the generic kernel lacks hardware-specific drivers and configurations, the rehosting environment must be carefully reconstructed to approximate the physical device's conditions by aligning kernel arguments, filesystem layouts, and device node populations with the original execution profile. This process unfolds across a multi-stage initialization workflow~\cite{chen2016firmadyne,kim2020firmae}, where each phase introduces distinct hurdles that impede full automation.

\textbf{Kernel Bootstrap.} This phase demands precise QEMU parameters, including CPU architecture, memory allocation, and kernel command-line arguments, alongside repair of root filesystem defects such as structural corruption or format mismatches. FirmAE mitigates some compatibility issues through runtime kernel arbitration, but remains reliant on architecture-default parameter templates. For firmware with vendor-specific dependencies or non-standard boot sequences, generic templates readily trigger rootfs mounting failures or kernel panics.

\textbf{User-space Initialization.} User-space boot is driven by \texttt{/etc} scripts (e.g., \texttt{rcS, init}) encoding network setup, service dependencies, and hardware checks. FirmAE~\cite{kim2020firmae} heuristically locates and executes the original init scripts, appending its own helper scripts at the end, thus preserving the full boot sequence. However, without any semantic analysis or filtering, it treats hardware‑related failures and legitimate service commands equally, thereby conflating essential service logic with redundant hardware probes and leading to deadlocks, termination, or incomplete initialization.

\textbf{Service Runtime.} Once initialization completes, network services (e.g., \texttt{httpd}, \texttt{sshd}) face silent crashes, segmentation faults, and infinite polling loops due to unemulated peripherals. FirmAE distills common failure patterns from 437 firmware images into five predefined arbitration strategies. While effective for known failure modes, this static, rule-bound approach cannot adapt to novel crashes or cascading dependencies. Consequently, runtime failures halt rehosting, forcing manual log inspection and a mix of dynamic and static analysis until the service becomes interactive.

\subsection{LLM-Assisted Firmware Analysis}
Recent advances in  LLMs have demonstrated strong code
comprehension and reasoning abilities~\cite{hou2024large,luo2023wizardcoder,roziere2023code}. The emergence of tool-use
mechanisms~\cite{shen2024llmtool,qin2024tool,hou2025model}, external knowledge
bases~\cite{lewis2020rag}, and LLM-based autonomous
agents~\cite{jin2024llms,wang2024survey} has further expanded the potential to automate complex
system-level tasks. In the context of firmware rehosting,
FlexEmu~\cite{lei2025flexemu} leverages LLMs to model heterogeneous MCU
peripherals by extracting semantic details from source code, reducing
manual effort in peripheral emulation. These developments indicate that LLMs
show growing promise for automating firmware rehosting tasks, though existing
work remains focused on specific stages such as peripheral modeling.

\begin{figure*}[h]
  \centering
  \includegraphics[width=\linewidth]{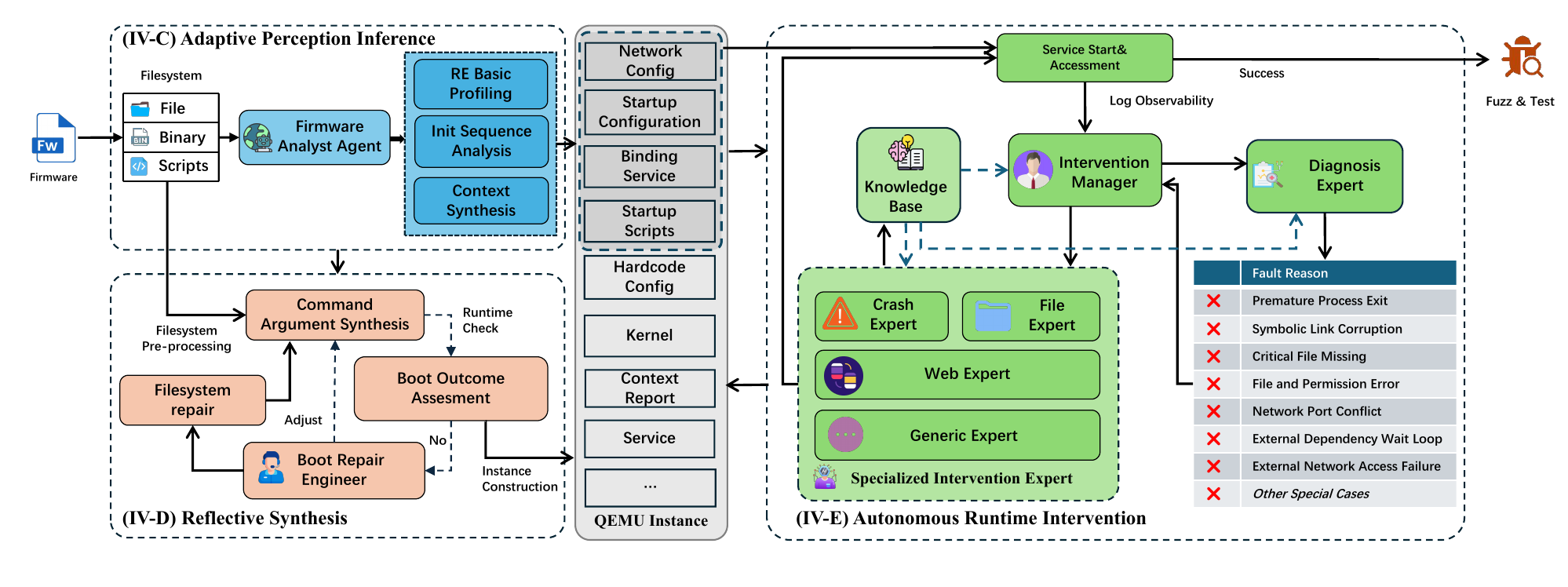}
  \caption{Overview of the \toolname}
  \label{fig:overview}
\end{figure*}

\section{Motivation}

Existing full-system rehosting frameworks (e.g., Firmadyne\cite{chen2016firmadyne}, FirmAE\cite{kim2020firmae}) address rehosting environmental reconstruction challenges through static arbitration mechanisms based on empirically derived rules. While effective for well-studied datasets, these approaches exhibit fundamental limitations when encountering diverse, unseen firmware images with complex environmental dependencies. The empirically derived rules lack generalization capability, rendering them ineffective against vendor-specific initialization logic or non-standard hardware configurations. Consequently, successful rehosting still relies heavily on human expertise to manually identify QEMU parameters, patch initialization scripts, and diagnose runtime crashes. These approaches lack the capability to semantically understand initialization logic, iteratively refine boot parameters based on execution feedback, or autonomously remediate runtime errors.

\textbf{Empirical Observations.} Our large-scale analysis of over 100 firmware images reveals that rehosting failures are highly heterogeneous and systematically cascade across the entire boot lifecycle. As summarized in Table~\ref{tab:failure_cases}, these obstacles manifest as distinct, stage-specific barriers that collectively stall the rehosting pipeline, which cannot be addressed by existing Approaches~\cite{kim2020firmae,tay2023greenhouse,qin2026firmwell}. At the \textbf{Kernel Bootstrap} and \textbf{Rootfs Mounting} stages, fundamental environment mismatches trigger immediate breakdowns. For the WAVLINK NU516U1, a kernel panic is caused by an unspecified CPU architecture during QEMU launch, and an feasible fix strategy is to explicitly adjust the \texttt{-cpu 74Kf} parameter. Concurrently, the D‑Link DGL-5500 fails to mount its root filesystem due to corruption and dangling symbolic links. These issues can be resolved by pre‑boot repair routines that remove broken links and restore missing shared libraries. During \textbf{User‑space Initialization}, boot scripts frequently encounter unresolvable hardware dependencies. The TRENDnet TEW‑813DRU enters an initialization deadlock when the \texttt{rcS} script awaits unresolved hardware configurations, and a feasible repair strategy is to inject minimal synthetic \texttt{.cfg} files to decouple dependency checks and resume execution. Finally, at the \textbf{Service Runtime} stage, services face structural and environmental mismatches.The Totolink NR1800X crashes due to \texttt{lighttpd} configuration path mismatches, which can be addressed by redirecting startup parameters to the correct path or creating a symlink. The Tenda AC18 experiences premature daemon termination when the \texttt{ConnectCfm} routine invokes \texttt{cfmd} under strict hardware bindings, a problem that can be bypassed by inserting \texttt{gdb} breakpoints or dynamically patching instructions. And the TP‑Link TL‑IPC43AN remains trapped in an infinite polling loop waiting for a missing \texttt{/tmp/jffs2\_ready} flag, where a straightforward fix is to dynamically generate the required placeholder file.

Collectively, these cases demonstrate that rehosting obstacles are fundamentally multi-dimensional and context-dependent, spanning structural inconsistencies, fragile boot parameters, and unpredictable runtime states. Static arbitration mechanisms inherently fail to address this spectrum, as their fixed heuristics cannot adapt to cross-stage dependencies or dynamically synthesize phase-specific repairs. This systemic fragmentation highlights the need for a closed‑loop, autonomous rehosting pipeline that can dynamically handle these obstacles. To this end, we propose \toolname, an autonomous and adaptive framework that handles these processes and obstacles, as detailed in Section~\ref{sec:design}.

\section{Design}
\label{sec:design}
In this section, we present the design of \toolname. We first define the problem scope (Section~\ref{sec:problem_scope}) and provide a system overview (Section~\ref{sec:system_overview}), followed by a sequential description of the Adaptive Perception Inference module (Section~\ref{sec:adaptive_perception}), the Reflective Synthesis module (Section~\ref{sec:reflective_synthesis}), and the Autonomous Runtime Intervention module (Section~\ref{sec:runtime_intervention}).

\subsection{Problem Scope} \label{sec:problem_scope}

\toolname targets \textbf{Linux-based embedded firmware} on common IoT architectures via QEMU full-system rehosting. Our goal is to \textbf{ensure network service interactivity} to support downstream security tasks, not just availability. We operate on the premise that firmware images are obtainable from public online repositories, vendor download portals, or through direct extraction from physical IoT devices. These acquired images supply the root filesystem required to construct a functional rehosting environment. This work focuses on vulnerability discovery via dynamic analysis. Success is defined by network services such as \texttt{httpd} and \texttt{sshd} starting, listening on expected ports, and responding to requests.

\subsection{System Overview}
\label{sec:system_overview}

\toolname operates as an LLM-guided, closed-loop firmware rehosting framework structured around three sequential phases. As illustrated in Figure~\ref{fig:overview}, the architecture orchestrates a continuous pipeline that transforms raw firmware binaries into stable, interactive rehosting environments. Each phase implements a distinct mechanism to address a specific stage of the rehosting lifecycle, progressively improving rehosting metrics.

\textbf{Adaptive Perception Inference (A.P.I)} (§\ref{sec:adaptive_perception}) employs an LLM-driven multi-stage analysis pipeline to autonomously parse firmware rootfs, extract architectural and service semantics, and classify components into hardware-dependent and service-critical categories. By treating firmware exploration as a goal-directed reasoning process, this phase resolves initialization decoupling challenges and synthesizes a structured context blueprint that grounds downstream operations in accurate firmware semantics.

\textbf{Reflective Synthesis (R.S)} (§\ref{sec:reflective_synthesis}) translates the extracted context into a bootable environment through filesystem pre-repair and parameter synthesis. A synthesis agent drives a validation loop that watches the boot progress, identifies failures from the output, and applies targeted fixes to launch arguments or filesystem structures. After each repair, the system repackages the environment before the next boot attempt. This process stabilizes kernel boot loading and rootfs mounting across heterogeneous architectures.

\textbf{Autonomous Runtime Intervention (A.R.I)} (\S\ref{sec:runtime_intervention}) deploys a hierarchical multi-agent framework that observes post-boot execution states and dynamically routes runtime faults to specialized domain experts. Through targeted binary bypass, environment remediation, and network configuration adaptation, this phase systematically resolves post-boot service blockers. By iteratively applying expert-driven repairs, the module progressively improves the rehosting metrics until full service interactivity is established.


Together, these three phases form a closed-loop rehosting pipeline unified by a \textit{cross-phase context cascade}. All agents share a common semantic context across the pipeline. This context helps agents track state, accumulate useful knowledge, and decide what to do next. Each phase uses it to guide repairs. All analytical and synthetic operations ultimately converge on a centralized \textit{QEMU Instance}, which serves as the persistent execution substrate for orchestrating initialization, stabilizing boot sequences, and validating runtime service responsiveness.

\subsection{Adaptive Perception Inference}
\label{sec:adaptive_perception}

The Adaptive Perception Inference module takes the extracted firmware rootfs as input and produces a Structured System Context Object. This object consolidates CPU architecture, web service entry points, initialization dependency maps, and validated boot configurations. At the core of this module is a \textit{Firmware Analyst Agent}, whose objective is to produce a minimal, hardware‑decoupled execution context that prevents early‑boot deadlocks while retaining full dependency information. This allows the Autonomous Runtime Intervention stage to selectively restore suppressed services in a stable environment. The module is achieved through a three‑stage, LLM‑guided analysis pipeline that progressively decouples hardware-specific probing from essential service logic.

The pipeline begins with \textbf{reverse engineering and basic profiling}, where the agent extracts foundational attributes by inspecting ELF headers, filesystem layouts, and binary strings. It determines CPU architecture, endianness, and C library variant. It also locates the primary httpd binary and extracts its compile‑time parameters such as web root, port, and CGI paths. These attributes collectively establish the baseline context for all downstream analysis tasks.

Building on this baseline, the agent proceeds to \textbf{init sequence analysis and hardware dependency decoupling}. The design of this component addresses a central tension in boot sequence reconstruction, since the init system must be simplified enough to run inside QEMU, yet no critical service dependency can be severed. To navigate this, we design a two-phase analysis pipeline. The agent first reconstructs the full boot sequence by parsing the init system (\texttt{rcS}, \texttt{inittab}, \texttt{profile}), tracing the exact chain of script invocations and services launches. Hardware dependency classification then proceeds through two complementary mechanisms. The first is a set of prompt constraints encoding domain knowledge. The agent flags processes known to deadlock or crash inside QEMU, such as vendor-specific hardware daemons, watchdog timers, and wireless-related processes. The second mechanism relies on programmatic binary analysis. The agent inspects symbol tables for hardware-facing calls, scans for system-call patterns that probe absent device nodes, and matches log strings against known hardware-initialization idioms.

Inspired by Housefuzz~\cite{xiao2025housefuzz}, a deliberate design choice concerns binaries that are tightly coupled with service-critical processes yet also carry hardware-probe logic, such as \texttt{system\_manager} and \texttt{ncc}. Simple removal would sever essential IPC channels or configuration delivery paths. We therefore add a deeper reverse-analysis step. The agent traces how the binary interacts with the target service daemon, identifies the IPC mechanisms that deliver configuration or control commands, and reconstructs the communication protocol. The output is a semantic dependency summary that records three things. First, it notes what the binary provides to the service. Second, it specifies how the binary expects to be launched. Third, it details what minimal logic can replace it without hardware probing. This summary, together with the hardware-dependency manifest, is preserved in the context object and handed to later stages to inform both boot synthesis and runtime intervention.

Next, the pipeline moves to \textbf{context synthesis}, where the agent generates a \textit{minimal startup configuration} that achieves service interactivity with zero hardware probing. It rewrites the initialization script to skip hardware‑dependent sections. It then validates that service‑critical daemons such as the HTTP server and DNS proxy still receive their required IPC, network bindings, and configuration files. The output is a verified startup script and an environment profile that guarantee a deadlock‑free transition from kernel handoff to user‑space service launch. The suppression manifest is preserved alongside this script, documenting exactly which services were excluded and which hardware events they were waiting for. This creates a direct regression pathway. The Autonomous Runtime Intervention module can later consult the manifest and re‑insert a previously suppressed service once the runtime environment gains the necessary virtual devices or network interfaces.

The pipeline concludes by assembling the \textit{Structured System Context Object}, which bundles the architectural profile, the minimal startup script, the environment configuration, and the suppression manifest. This design achieves two critical goals. First, it gives the R.S module with a clean, hardware‑agnostic boot blueprint that dramatically reduces the search space for kernel parameter tuning. Second, it equips the A.R.I modules with complete semantic knowledge of previously excluded components, enabling it to autonomously re‑integrate services when the runtime environment becomes capable of handling them. In this way, \toolname enables a \textit{progressive rehosting} strategy. First launch the minimum set of services needed to establish basic connectivity, then selectively restore additional components based on runtime error feedback.

\subsection{Reflective Synthesis}
\label{sec:reflective_synthesis}

The Reflective Synthesis module translates the Structured System Context Object into a bootable, user‑space‑ready QEMU instance through a closed‑loop refinement process. At the core of this module is a \textit{Boot Repair Engineer Agent}. Static mapping of extracted architecture, memory, and kernel parameters cannot fully anticipate the precise combinations required by the target firmware. Mismatched CPU flags, insufficient memory, or incompatible kernel versions often surface only at runtime. Thus, the module does not rely on a single static configuration. Instead, it synthesizes an initial boot hypothesis and launches QEMU. If the boot fails, an LLM-guided agent assesses the outcome and iteratively adjusts the QEMU command line and root filesystem until a stable user-space prompt appears or the retry limit is reached.

\textbf{Filesystem Pre-processing.} 
Prior to the first boot attempt, the module performs targeted, dependency‑aware filesystem remediation guided by results of the A.P.I module. Drawing on practices established in prior work~\cite{kim2020firmae,tay2023greenhouse}, we preemptively resolve common structural and dependency faults to prevent excessive intervention during subsequent refinement:  
\begin{itemize}
\item Inject missing or malformed system configuration files, such as \texttt{/etc/passwd} and \texttt{/etc/group}.
\item Deploy NVRAM virtualization libraries when proprietary hardware calls are detected, following Greenhouse~\cite{tay2023greenhouse}.
\item Reconstruct broken symbolic links.
\item Correct execute permissions on core binaries and dynamic linkers.
\item Package the hardware‑decoupled startup scripts from the A.P.I module, along with service launch and network configuration scripts, into the filesystem at standard entry points.
\end{itemize}
By analyzing a set of hook-failure cases, we found that conventional environment-based hooking is often unreliable on stripped uClibc/musl firmware, where \texttt{LD\_PRELOAD} support may be disabled and section metadata may be unavailable. To address this limitation, we complement these filesystem repairs with a dual-path library interposition strategy. When the target loader supports \texttt{LD\_PRELOAD}, we use conventional environment-based injection. Otherwise, we inject a \texttt{DT\_NEEDED} entry directly into the target binary through our ELF \texttt{DYNAMIC} segment patcher, allowing the same compatibility shim to be loaded even when preload-based interposition is not usable.

These operations are orchestrated as a non‑destructive process. Dependency‑aware modifications preserve the original filesystem structure while surgically neutralizing hardware‑specific initialization blockers. By resolving structural and dependency faults before the first boot attempt, the filesystem significantly reduces early‑stage failures and narrows the configuration adjustments required during subsequent iterative refinement.

\textbf{Command Argument Synthesis and Boot Outcome Assessment.} Rather than enumerating a large space of possible QEMU configurations, the module synthesizes the initial boot command directly from the Structured System Context Object produced by A.P.I module. The extracted architecture, memory layout, and kernel entry point provide a grounded starting estimate. This rules out broad categories of invalid combinations and forms a high‑confidence hypothesis, so that subsequent refinement only needs to correct a small number of residual mismatches. At each boot attempt, the module monitors the QEMU serial console output for a single success signal, namely the appearance of a stable user‑space shell prompt that provides an unambiguous pass and fail boundary. When the prompt appears, the kernel has successfully mounted the root filesystem, executed the init sequence, and transitioned to user‑space execution. when boot fails, the module captures the full console trace and extracts the failure signature (kernel panic, mount error, or init crash) into a structured diagnostic record that matches the failure categories in the agent's knowledge base, enabling rapid, targeted repair without parsing raw logs.

\textbf{Agent-Driven Iterative Refinement.} 
The \textit{Boot Repair Engineer Agent} is built around two design choices. First, the agent is constrained to propose exactly one targeted modification per iteration. A single adjustment may be a QEMU parameter change or a filesystem patch, but never both simultaneously. Without this constraint, the agent tends to bundle multiple fixes into one response, making it impossible to attribute success or failure to a specific change. This constraint serves two purposes. First, it makes the effect of each action clearly attributable, keeping the search trajectory tractable. Second, it prevents over-correction that could destabilize a working configuration. Every iteration step is recorded as a discrete node with clear inputs and outputs, along with a success or failure label, making the entire refinement history auditable and reproducible.

Second, the agent does not reason from scratch when a boot fails. The LLM often drifts into open‑ended reasoning loops or proposes plausible but incorrect fixes based on general Linux knowledge rather than firmware‑specific patterns. To ground the agent, we provide a knowledge base that maps failure signatures to their most probable root causes and corresponding repair actions. The agent uses this mapping to form an initial diagnostic hypothesis, then invokes dedicated verification tools to confirm the hypothesis before executing the planned adjustment. Only after tool-based confirmation does the agent apply the single modification. This diagnose-then-act workflow prevents wasteful replacement of functional components and limits the loop to a maximum of five iterations.

The Agent handles two fault categories, distinguished by when they occur. Failures that surface before the kernel reaches user‑space typically originate from virtual hardware misconfiguration, for example an \texttt{Illegal instruction} crash due to a mismatched CPU model or a kernel decompression panic from insufficient memory. For these, the agent performs QEMU Parameter Rectification. A concrete case is WAVLINK NU516U1, which triggers a kernel panic early in boot. The agent inspects the console log, cross‑references the panic signature with the firmware's architectural metadata supplied by the A.P.I module, and invokes its analysis tools to determine that the required \texttt{-cpu} parameter is missing. It then adjusts the QEMU command line to include \texttt{-cpu 74Kf} and reboots, after which the kernel panic disappears. Failures after kernel handoff, where processes fail to spawn or terminate unexpectedly, instead require Filesystem Remediation, which includes injecting missing shared libraries from a compatibility pool, resolving broken symbolic links, and correcting permission masks. All modifications are batched into a fresh disk image before the next boot attempt.

Every intervention is appended to a structured, persistent repair history. Before proposing any new action, the agent explicitly checks this history to avoid repeating previously attempted fixes. This simple yet effective memoization mechanism prevents cyclic loops where the agent would otherwise oscillate between two conflicting adjustments. By coupling emulator configuration with rootfs integrity under LLM-guided, evidence-driven reasoning, this reflection mechanism systematically eliminates boot-time blockers without manual intervention, ensuring that only fully initialized, user-space-ready instances are handed off to the subsequent Autonomous Runtime Intervention module.

\subsection{Autonomous Runtime Intervention}
\label{sec:runtime_intervention}

Once the R.S module hands off a booted, user-space-ready QEMU instance, the Autonomous Runtime Intervention module takes over. It configures the virtual network interface and applies the minimal startup script produced by the A.P.I module, which launches only the service-critical daemons. After the script is applied, the module employs a multi-agent system to monitor the runtime state and advance the rehosting metrics in stages.

\textbf{Multi-Agent Design.}
Runtime failures at this stage, such as crashes, missing files, unbound ports, or misconfigured web services, rarely occur in isolation. Resolving one fault often exposes another, and the required diagnostics span multiple domains that a single LLM agent struggles to cover reliably. Moreover, firmware analysis involves extensive context, such as kernel logs, filesystem layouts, and binary disassembly, and a diverse set of tools spanning dynamic analysis, static analysis, and network validation. A monolithic agent, when faced with this overloaded context and broad tool set, tends to hallucinate diagnoses or apply overly broad fixes that disrupt previously working components. By partitioning responsibilities across a Manager Agent and domain-specific specialists, each agent operates within a focused context and a limited, relevant toolchain. This effectively mitigates context explosion and tool misuse. We adopt a multi-agent design in which a Manager Agent observes and routes faults, specialist agents repair within narrow domains, and the Manager coordinates sequential handoffs to resolve compound failures and progressively advance the rehosting metrics.

\textbf{Fault Observation and Routing.}
A \textit{Manager Agent} sits at the center of the architecture, acting solely as an observer and router. This design strictly separates diagnosis from repair. In a single-agent setup, the LLM that diagnoses a fault also tries to fix it, often applying premature or incorrect repairs that corrupt the system. By keeping the Manager read-only, a misdiagnosis never damages the running instance, and specialists always receive a clean, consistent snapshot. The Manager continuously monitors the console output and process list for failure signatures such as repeated errors, persistent crash logs, or ports that remain unbound despite a launched service. When a fault is detected, the Manager classifies it by system layer and failure category. Failures matching known patterns with high confidence are routed directly to the appropriate specialist. For ambiguous or compound failures that cannot be confidently classified, the Manager delegates the task to a dedicated \textit{Diagnosis Agent}. This agent performs deeper analysis to disambiguate the root cause. Once the fault is clearly identified, the \textit{Diagnosis Agent} returns the result to the Manager, which then routes it to the correct specialist. Throughout this process, the Manager never executes repairs itself, so a misdiagnosis cannot compromise the system.


\textbf{Specialist Intervention Agents.}
We distribute remediation across specialist agents rather than a single general‑purpose LLM. A monolithic agent, given all tools at once, often picks the wrong diagnostic path and misses domain‑specific failure patterns. Scoping each specialist to one fault category with a focused toolchain keeps diagnoses grounded and repairs precise. The \textit{CrashExpert} handles services that terminate early due to compulsory hardware presence checks. It first statically locates blocking code paths, then employs LLM‑guided semantic understanding of the decompiled code to determine precisely where to insert breakpoints. Instead of applying binary patches~\cite{tay2023greenhouse}
, we use breakpoint‑chain intervention to bypass checks at runtime without modifying the binary. This eliminates the danger of an LLM agent introducing hard‑to‑detect damage into the firmware image. The \textit{FileExpert} resolves runtime filesystem faults such as missing configuration files, broken symbolic links, or incorrect permissions. It correlates process access traces with the rootfs structure and consults binary metadata to infer expected file formats when surface‑level errors are insufficient. The \textit{WebExpert} bridges the gap from port availability to full HTTP interactivity. It follows an ordered diagnostic sequence that verifies web root permissions, validates CGI interpreter paths, and extracts hard‑coded compile‑time paths from the httpd binary only when earlier checks pass.

Failures that fall outside these three well‑defined categories are escalated to the \textit{GenericExpert}. Unlike the narrow specialists, the \textit{GenericExpert} has unrestricted access to all tools and knowledge sources. This mirrors the monolithic agent discussed earlier, but with a critical difference. It is invoked only as a fallback when targeted repairs fail, not as the default for every fault. Its primary role is exploratory diagnosis for previously unseen or multi‑domain failures. When no existing pattern matches, the \textit{GenericExpert} investigates across layers, formulates a repair hypothesis. If successful, it will record the new pattern so that similar faults can be handled by the appropriate specialist without repeating costly exploration.

\textbf{Sequential Fault Handoff.}
This sequential handoff is essential for resolving compound failures. A single agent, when confronted with multiple overlapping faults, often attempts to fix them all at once and loses track of cause and effect. By contrast, our pipeline addresses one fault at a time and re-evaluates the system state after each repair, so that newly exposed failures are correctly attributed and routed. For example, the httpd daemon first crashes due to a hardware‑probe check, which \textit{CrashExpert} bypasses via GDB breakpoints. Once bypassed, the same daemon fails to start because a configuration file was never provisioned, which \textit{FileExpert} resolves. Even after both fixes, the web service returns errors from mismatched compile‑time paths, which \textit{WebExpert} corrects. The \textit{Manager's} routing loop dispatches each emergent fault to the correct specialist and iterates until all symptoms are cleared. This design decomposes compound failures into a sequence of targeted, single‑domain repairs validated at each step.

\textbf{Knowledge Reuse Across Firmware.}
Without reuse, the agent treats every fault as novel and invokes the full LLM diagnostic pipeline from scratch, which is both slow and costly. To avoid this, every validated repair trajectory records the fault signature, diagnostic steps, specialist invoked, and applied fix in a structured knowledge base indexed by failure category and architectural context. When the \textit{Manager Agent} encounters a fault on a subsequent firmware image, it first queries this knowledge base for signatures that match within a similarity threshold. A successful match allows the corresponding specialist to apply the proven fix directly after a lightweight verification step, bypassing the full diagnostic pipeline. Unsuccessful matches still benefit from partial knowledge. The Manager can narrow the routing decision by comparing the new fault against previously catalogued patterns even when no exact match exists. Over a sequence of firmware images, this feedback loop accumulates a growing library of reusable solutions, progressively reducing the average intervention cost per firmware.

\section{Implementation}
\label{sec:implementation}

Our prototype implementation of \toolname comprises roughly 16,000 lines of code, primarily written in Python, and is built on top of CrewAI~\cite{crewai}, an open-source multi-agent orchestration framework. We leverage CrewAI's \texttt{Flow} module to coordinate the three-phase pipeline as an event-driven state machine.

\textbf{Multi-agent architecture.}
\toolname organizes agents into phase-specific crews. A.P.I employs a single analyst agent that sequentially executes subtasks to extract firmware primitives. The R.S module employs an LLM-driven reflective synthesis process, in which the agent proposes QEMU configurations and filesystem repair, validates them against boot telemetry, and iteratively refines the configuration on failure. In Phase 3, we use a hierarchical arrangement. A manager agent first diagnoses the fault. It then hands the fault over to one of four specialist agents, namely CrashExpert, FileExpert, WebExpert, or GenericExpert. Each specialist is equipped with a carefully chosen set of allowed tools and a knowledge base tailored to its domain.

\textbf{Prompt and knowledge design.}
Each agent in \toolname is guided by carefully crafted prompts that define its role, reasoning strategy, and expected output format, steering the LLM to accomplish domain-specific tasks reliably. Beyond per-agent prompt engineering, \toolname decouples domain knowledge from agent logic via JSON-structured knowledge schemas. Each expert is associated with a knowledge file defining fault classification rules, repair workflows, and domain heuristics, which are injected at runtime through CrewAI's \texttt{StringKnowledgeSource}. Task prompts enforce structured JSON output, enabling the flow controller to track each intervention action.

\textbf{Tool design.}
\toolname provides seven specialized tools following a unified pattern in which a backend class encapsulates implementation logic and a CrewAI \texttt{BaseTool} wrapper exposes it to agents. The tools cover static analysis powered by radare2~\cite{radare2}, dynamic debugging via GDB remote protocol, disk image manipulation on ext4 through \texttt{debugfs}, QEMU lifecycle management, file system operations, network testing, and multi-tier validation. Tools are filtered per expert via an \texttt{EXPERT\_TOOL\_MAP}.

\textbf{Backend.}
\toolname communicates with LLMs through the LiteLLM~\cite{litellm_docs_en} gateway, which normalizes provider-specific APIs into a unified interface. It currently supports QEMU~\cite{bellard2005qemu} emulation targeting ARM and MIPS variants.

\section{Evaluation}
To evaluate the effectiveness of \toolname, we conducted a comprehensive set of experiments designed to address the following research questions:

\begin{itemize}
    \item \textbf{RQ1:} How does \toolname compare to existing state-of-the-art tools across key rehosting metrics?
    \item \textbf{RQ2:} How effectively do \toolname's specialized components generalize to diverse firmware rehosting obstacles across heterogeneous targets?
    \item \textbf{RQ3:} How effectively does \toolname enable high-fidelity vulnerability reproduction and novel security analysis?
    \item \textbf{RQ4:} How do \toolname's token consumption and execution latency vary across heterogeneous firmware samples?
\end{itemize}
\subsection{Evaluation Setup}

\begin{table*}[t]
\centering
\caption{Firmware Rehosting Results across FirmAE, Greenhouse, Firmwell and \toolname}
\label{tab:firmware_results}
\footnotesize
\resizebox{\textwidth}{!}{%
\begin{tabular}{c|c|l|c|
    *{4}{c}|   
    *{3}{c}|   
    *{3}{c}|   
    *{4}{c}}   

\toprule
\multirow{2}{*}{\textbf{No}} & \multirow{2}{*}{\textbf{Vendor}} & \multirow{2}{*}{\textbf{Device}} & \multirow{2}{*}{\textbf{Arch}} &
\multicolumn{4}{c|}{\textbf{FirmAE}} &
\multicolumn{3}{c|}{\textbf{Greenhouse}} &
\multicolumn{3}{c|}{\textbf{Firmwell}} &
\multicolumn{4}{c}{\textbf{\toolname}} \\
\cmidrule(lr){5-8} \cmidrule(lr){9-11} \cmidrule(lr){12-14} \cmidrule(lr){15-18}
& & & &
\textbf{Ker.} & \textbf{Net.} & \textbf{Port} & \textbf{Inter.} &
\textbf{Exe.} &  \textbf{Port} & \textbf{Inter.} &
\textbf{Exe.} &  \textbf{Port} & \textbf{Inter.} &
\textbf{Ker.} & \textbf{Net.} & \textbf{Port} & \textbf{Inter.} \\
\midrule
1 & \multirow{2}{*}{Totolink} & NR1800X & mipsel & \cmark & \cmark & \xmark & \xmark & \cmark & \cmark & \cmark & \cmark & \cmark & \cmark & \cmark & \cmark & \cmark & \cmark\\
2 & & N150RT & mipsel & \cmark & \cmark & \cmark & \cmark & \cmark & \xmark & \xmark & \cmark & \xmark & \xmark & \cmark & \cmark & \cmark & \cmark \\
\midrule
3 & \multirow{4}{*}{D-link} & DAP-1522 & mipsel & \cmark & \cmark & \cmark & \cmark & \cmark  & \xmark & \xmark & \cmark & \xmark & \xmark & \cmark & \cmark & \cmark & \cmark \\
4 & & dsp-w215 & mipsel & \cmark & \cmark & \cmark & \xmark & \cmark & \xmark & \xmark & \cmark & \xmark & \xmark & \cmark & \cmark & \cmark & \cmark \\
5 & & DGL-5500 & mipsel & \xmark & \xmark & \xmark & \xmark & \cmark & \xmark & \xmark & \cmark & \xmark & \xmark & \cmark & \cmark & \cmark & \cmark  \\
6 & & DIR823X & arm64 & \xmark & \xmark & \xmark & \xmark & \xmark & \xmark & \xmark & \xmark & \xmark & \xmark &  \cmark & \cmark & \cmark & \cmark \\
\midrule
7 & \multirow{2}{*}{WAVLINK} & NU516U1 & mipsel & \xmark & \xmark & \xmark & \xmark & \xmark & \xmark & \xmark & \xmark & \xmark &\xmark  & \cmark & \cmark & \cmark & \cmark  \\
8 & & WN531P3 & mipsel & \cmark & \cmark & \xmark & \xmark & \cmark & \xmark & \xmark & \cmark & \xmark & \xmark & \cmark & \cmark & \cmark & \cmark  \\
\midrule
9 & \multirow{4}{*}{Tenda} & AC15 & armhf & \cmark & \cmark & \xmark & \xmark & \cmark &   \xmark & \xmark & \cmark & \xmark & \xmark & \cmark & \cmark & \cmark & \cmark  \\
10 & & AC18 & armhf & \cmark & \cmark & \xmark & \xmark & \cmark &  \xmark & \xmark & \cmark & \xmark & \xmark & \cmark & \cmark & \cmark & \cmark   \\
11 & & AC500 & armhf & \cmark & \cmark & \xmark & \xmark & \cmark &  \xmark & \xmark & \cmark & \xmark & \xmark & \cmark & \cmark & \cmark & \cmark   \\
12 & & AX1806 & armhf & \cmark & \cmark & \xmark & \xmark & \cmark &  \xmark & \xmark & \cmark & \xmark & \xmark & \cmark & \cmark & \cmark & \cmark  \\
\midrule
13 & \multirow{1}{*}{Draytek}& Vigor3900 & armel & \xmark & \xmark & \xmark & \xmark & \xmark & \xmark & \xmark & \xmark & \xmark & \xmark & \cmark & \cmark & \cmark & \cmark \\
\midrule
14 & \multirow{2}{*}{Netgear} & wn1000rp & mipsel & \cmark & \cmark & \xmark & \xmark & \cmark & \cmark & \cmark & \cmark & \cmark & \cmark & \cmark & \cmark & \cmark & \cmark   \\
15 & & XR500 & armhf & \cmark & \xmark & \xmark & \xmark & \cmark & \xmark & \xmark & \cmark & \xmark & \xmark & \cmark & \cmark & \cmark & \xmark   \\
\midrule
16 & \multirow{2}{*}{TRENDnet} & TEW-711BR & mipseb & \cmark & \cmark & \xmark & \xmark & \cmark & \xmark & \xmark &  \xmark & \xmark & \xmark & \cmark & \cmark & \cmark & \cmark  \\

17 &  & TEW-813DRU & mipseb & \cmark & \cmark & \xmark & \xmark & \xmark & \xmark & \xmark & \xmark & \xmark & \xmark & \cmark & \cmark & \cmark & \cmark  \\
\midrule
18 & \multirow{2}{*}{TP-LINK} & TL-IPC43AN & armhf & \cmark & \cmark & \xmark & \xmark & \xmark & \xmark & \xmark & \xmark & \xmark & \xmark & \cmark & \cmark & \cmark & \cmark \\
19 & & RE580D & armel & \cmark & \cmark & \xmark & \xmark & \xmark & \xmark & \xmark & \xmark & \xmark & \xmark & \cmark & \cmark & \cmark & \cmark \\
\midrule
20 & \multirow{1}{*}{Asus} & FW\_WL500gPv2\_2015 & mipsel & \cmark & \cmark & \xmark & \xmark & \cmark & \xmark & \xmark & \cmark & \cmark & \cmark & \cmark & \cmark & \cmark & \cmark  \\
\midrule
21 & \multirow{1}{*}{XIAOMI} & AX9000 & arm64 & \xmark & \xmark & \xmark & \xmark & \xmark & \xmark & \xmark & \xmark & \xmark & \xmark & \cmark & \cmark & \cmark & \xmark  \\
\midrule
\multicolumn{4}{c|}{\textbf{Total}} & 
76.2\% & 71.4\% & 14.3\% & 9.5\% & 
66.7\% & 9.5\% & 9.5\% & 
61.9\% & 19.0\% & 19.0\% & 
100.0\% & 100.0\% & 100.0\% & 90.5\% \\
\bottomrule
\end{tabular}%
} 
\end{table*}

\textbf{Dataset.} 
To evaluate \toolname under heterogeneous conditions, we use a deliberately diverse set of \textbf{21 representative samples} drawn from 10 major vendors. These samples span \textbf{five diverse architectures} and cover multiple device types such as routers, cameras, and IoT devices. This heterogeneous composition ensures a robust test of \toolname's adaptability while allowing detailed manual verification of rehosting fidelity.

\textbf{Baselines.} We compare \toolname against three state-of-the-art frameworks representing distinct technical paradigms. FirmAE~\cite{kim2020firmae} stands as a leading approach in full-system rehosting by introducing arbitrated emulation, which applies targeted heuristic fixes at specific arbitration points during firmware boot to maximize the launch of user‑space services rather than faithfully emulating all hardware. User-space approaches \textbf{Greenhouse}~\cite{tay2023greenhouse} and \textbf{Firmwell}~\cite{qin2026firmwell} represent the state-of-the-art in user-mode rehosting and dependency-aware rehosting. Greenhouse pioneers dynamic API virtualization by mapping hardware dependencies to eight barrier types and applying ten predefined interception templates, while Firmwell advances rehosting by leveraging native firmware initialization routines to materialize runtime dependencies within a coordinated multi-process environment, augmented by dependency profiling and automated error inference.

\textbf{Environment.} All experiments were conducted on a server cluster equipped with Intel(R) Xeon(R) Gold 6248R CPU @ 3.00GHz and 320GB RAM. Each firmware sample was allotted a maximum execution time of 45 minutes.

\textbf{Rehosting Metrics.} 
To evaluate rehosting fidelity for security research, we adopt a hierarchical metric progression centered primarily on \textbf{web services}. Web interfaces constitute the most critical attack surface in IoT firmware and are highly sensitive to runtime environment correctness. We measure progression through four sequential stages that reflect increasing levels of service readiness, aligned with Table~\ref{tab:firmware_results}:

\begin{enumerate} 
    \item \textit{Kernel Boot (Ker.)}: The percentage of samples that successfully boot the kernel and mount the root filesystem. This indicates whether basic system initialization completes without kernel panics.
    
    \item \textit{Network Connectivity (Net.)}: The percentage of samples where the emulated network stack is functional and the device obtains a valid IP address. This verifies that virtualized network interfaces are properly initialized for external reachability.
    
    \item \textit{Service Port Availability (Port)}: The percentage of samples where the target web service successfully spawns and binds to its expected listening port. This confirms process readiness and accessibility for network scanning.
    
    \item \textit{Service Interactivity (Inter.)}: The percentage of samples capable of processing stateful HTTP requests (e.g., authentication, configuration retrieval, CGI execution). This represents the highest fidelity level, strictly required for downstream vulnerability analysis such as web fuzzing or exploit reproduction.
\end{enumerate}

This multi-stage design captures whether firmware progresses from basic execution to full service interactivity, ensuring it reaches a state viable for security analysis. Since \toolname operates as a full-system rehosting framework, it is evaluated across all four hierarchical stages to demonstrate end-to-end kernel-to-service readiness. In contrast, user-space baselines such as Greenhouse and Firmwell inherently bypass kernel initialization and virtual network stack provisioning. For these frameworks, we consolidate the initial boot and network phases into a unified \textbf{Execute} metric, which quantifies successful user-space service launch and runtime binding~\cite{qin2026firmwell,tay2023greenhouse}. We therefore evaluate each baseline strictly on the stages natively supported by its architectural paradigm, ensuring a fair and technically sound cross-paradigm comparison.

\begin{table*}[t]
\centering
\caption{Case Study: Diversity of Rehosting Obstacles and Targeted Interventions}
\label{tab:case_study}
\footnotesize 
\renewcommand{\arraystretch}{1.2} 
\setlength{\tabcolsep}{2pt} 
\begin{tabular}{@{} 
    c  
    >{\centering\arraybackslash}m{3cm} 
    >{\raggedright\arraybackslash}m{3cm} 
    >{\centering\arraybackslash}m{2.5cm} 
    >{\raggedright\arraybackslash}m{6.5cm} @{} 
}
\toprule
\textbf{ID} & \textbf{Device Model} & \textbf{Trigger / Cause} & \textbf{Repair Phase} & \textbf{Targeted Action} \\
\midrule
1 & \makecell{Tenda AC15} & \texttt{ConnectCfm} failure & (CrashExpert) A.R.I & GDB breakpoint \& alter register to bypass \\
\midrule
\multirow{4}{*}{2} 
& \multirow{4}{*}{\makecell{D-Link DGL-5500}} 
& \texttt{libm.so.0} missing & R.S & Search lib for injection \& replace symlink \\
& & Null symlink config & (FileExpert) A.R.I & Restore \texttt{lighttpd.conf} from \texttt{lighttpd\_base.conf} \\
& & IPv6 unsupported in VM & (FileExpert) A.R.I & \texttt{sed} disable \texttt{server.use-ipv6} \\
& & File descriptor limit & (FileExpert) A.R.I & \texttt{sed} cap \texttt{server.max-fds}=512, \texttt{max-connections}=256 \\
\midrule
\multirow{2}{*}{3} 
& \multirow{2}{*}{\makecell{TRENDnet TEW-711BR}} 
& Missing hardcoded \texttt{doc\_root} & (FileExpert) A.R.I & Symlink \texttt{/www} to hardcoded path \\
& & Missing PID directory & (FileExpert) A.R.I & \texttt{mkdir -p /var/run} \\
\midrule
4 & \makecell{WAVLINK NU516U1} & Kernel panic & R.S & Adjust QEMU \texttt{-cpu 74Kf} parameter \\
\midrule
5 & \makecell{Dlink DSP-W215} & Unextracted \texttt{www.tgz} & (WebExpert) A.R.I & Extract to specified directory /www \\
\midrule
6 & \makecell{WAVLINK WN531P3} & --- & --- & Direct Boot \\
\bottomrule
\end{tabular}
\end{table*}

\subsection{Firmware Rehosting Results}
\label{subsec:rq1}

We compare the rehosting performance of \toolname against FirmAE, Greenhouse, and Firmwell across 21 heterogeneous firmware images. The evaluation tracks rehosting progression in stages, from kernel or user-space execution through network initialization and port binding to final service interactivity. We further analyze the technical reasons behind the successes and failures across all firmware samples.

\textbf{Overall Results.} We evaluate \toolname against FirmAE, Greenhouse, and Firmwell across 21 firmware samples spanning nine vendors. As shown in Table~\ref{tab:firmware_results}, \toolname reaches a 100\% success rate at the Kernel, Network, and Port stages, and achieves a 90.5\% end-to-end interaction rate—the only tool to systematically advance beyond early execution stages across all architectures. In contrast, FirmAE, Greenhouse, and Firmwell stall at substantially lower interaction rates (9.5\%, 9.5\%, and 19.0\% respectively). The performance gap widens sharply at the Port stage. FirmAE exposes only 14.3\% of targets at the port level, while Greenhouse and Firmwell reach 9.5\% and 19.0\%. \toolname eliminates this bottleneck entirely, progressing all 21 samples to a listening port.

This stark disparity stems from concrete technical obstacles that static heuristics cannot overcome. On \texttt{WAVLINK NU516U1}, all three baselines fail because the firmware requires a specific CPU model and memory layout that their fixed rehosting profiles do not satisfy. \toolname detects the mismatch and adjusts the QEMU command accordingly. For \texttt{DIR823X}, an ARM64 firmware, FirmAE provides no ARM64 support, while Greenhouse and Firmwell, upon detecting ARM64, report an error and fail to identify the architecture. \toolname correctly extracts the ISA and boots it successfully. \texttt{TL-IPC43AN} hangs on a \texttt{/tmp/jffs2\_ready} wait loop, a file-system-level deadlock that no baseline resolves. Our analysis recognizes and removes the blocking condition. Tenda AC15 and AC18 contain a \texttt{ConnectCfm} hardware-presence check in their service binaries that permanently stalls user-space initialization under FirmAE, Greenhouse, and Firmwell. \toolname bypasses this check via breakpoint chaining. On \texttt{DAP-1522}, FirmAE achieves interaction by intercepting system calls to generate the required \texttt{rgdb} configuration, a capability that Greenhouse and Firmwell lack. \toolname reaches the same result through LLM-guided configuration synthesis. For \texttt{TEW-813DRU}, FirmAE opens the web port but fails to deliver interaction due to missing web-specific configuration files. \toolname advances to full interaction by inferring and synthesizing the required files.

The two remaining outliers, \texttt{XIAOMI AX9000} and \texttt{Netgear XR500}, represent newer and more complex firmware on which \toolname successfully advances to the Port stage through targeted intervention but cannot reach full interaction within the time budget. For \texttt{AX9000}, post-experiment analysis reveals a segmentation fault during HTTP request handling, rooted in deeply entangled dependencies across multiple proprietary daemons. The complexity of the failure pattern makes accurate root-cause identification infeasible within the bounded intervention cycles. For \texttt{XR500}, the obstacle centers on the \texttt{ubusd} daemon, which fails to start during user-space initialization. Without a functional \texttt{ubus} IPC bus, CGI scripts cannot communicate with backend services, leaving the web server stuck at port-open with no responsive content. Despite extended diagnostic effort, the agent could not resolve the \texttt{ubus} startup failure within the iteration budget, and this state persisted in follow-up experiments. These edge cases highlight a boundary condition where the most complex modern firmware demands longer diagnostic horizons or deeper architectural knowledge than the current time budget allows, pointing to a direction for future improvement.

\subsection{Intervention Strategies Across Diverse Obstacles}
\label{subsec:rq2}

To understand how \toolname handles diverse rehosting obstacles, we conducted a detailed analysis of the boot processes across all 21 firmware samples and identified three distinct levels at which our pipeline resolves failures. Table~\ref{tab:case_study}{} summarizes representative cases for each level.

\textbf{Level~1: Semantic Decoupling at the Perception Stage.}
The A.P.I module executes for every firmware image and produces a hardware-decoupled startup configuration that feeds into the subsequent runtime environment reconstruction. For a subset of firmware, this stage alone suffices to achieve functional rehosting without any subsequent intervention. In our evaluation, 6 out of the 21 samples boot directly to full service interactivity with no further repairs, as listed in Appendix~\ref{tab:case_study_ext}. A representative case is WAVLINK WN531P3. Under FirmAE, the device deadlocks due to vendor-specific hardware daemons probing absent physical devices. Our perception stage strips these processes while preserving service-critical daemons, allowing clean boot with zero downstream intervention. For these part of firmwares where hardware and software configurations are clearly separated during initialization, they can boot normally after the A.P.I stage.

\textbf{Level~2: Reflective Synthesis via Boot-Time Configuration and Filesystem Repair.}
A second class of failures surfaces during kernel bootstrap or filesystem mounting and must be resolved before user-space execution can begin. Only 3 of the 21 firmware samples require intervention at this stage, as listed in Appendix~\ref{tab:case_study_ext}. Case~4 (\texttt{WAVLINK NU516U1}) triggers a kernel panic due to a mismatched CPU model, which the Reflective Synthesis stage fixes by adjusting the QEMU \texttt{-cpu} parameter. Case~2 (D-Link DGL-5500) has a missing shared library \texttt{libm.so.0} that causes \texttt{/bin/sh} to fail, preventing the system from reaching user space. These cases confirm that failures at this stage, if left unresolved, permanently stall rehosting progress before Kernel Boot, making the Reflective Synthesis stage essential for advancing the pipeline.

\textbf{Level~3: Runtime Fault Resolution by Specialist Agents.}
Even after the first two stages have delivered a booted, user-space-ready instance, diverse runtime failures can still unexpectedly emerge. Of the 21 firmware samples, 12 require intervention at this stage, totaling 26 targeted repairs across four specialist agents, as detailed in Appendix~\ref{tab:case_study_ext}. Among them, the CrashExpert handles 4 cases involving process-level deadlocks, such as Case~1 (Tenda AC15), where the \texttt{ConnectCfm} hardware-presence check blocks the web service and is bypassed via GDB breakpoint chaining without modifying the binary. The FileExpert is the most frequently invoked, resolving 18 filesystem-level issues including null symlink configurations, IPv6 unavailability, file descriptor limits (Case~2, D-Link DGL-5500), and hardcoded path reconciliation (Case~3, TRENDnet TEW-711BR). The WebExpert addresses web-layer faults such as unextracted web roots (Case~5, Dlink DSP-W215), while the GenericExpert handles unclassified failures in 2 cases. These results show that the specialist agents collectively cover process, filesystem, web-service, and general fault domains, and their sequential orchestration under the Manager Agent enables systematic resolution of compound failures.

Taken together, these three levels form a progressive intervention spectrum. Level~1 handles cleanly separable hardware dependencies through semantic decoupling. Level~2 resolves boot-time blockers before Kernel Boot through Reflective Synthesis. Level~3 addresses diverse post-boot failures through coordinated specialist agents. Removing any single stage leaves its failure category unresolved, validating the staged architecture of \toolname.

\subsection{Vulnerability Reproduction and Discovery}
\label{subsec:rq3}

To assess the practical security value of \toolname, we evaluate its ability to reproduce known vulnerabilities and discover new ones in real-world firmware. Effective dynamic analysis requires a stable, interactive service runtime that survives initialization deadlocks and remains reachable over extended periods. \toolname achieves this by autonomously stripping hardware-dependent boot logic and intervening at runtime to keep critical services alive. We apply the framework to a set of router firmware images and perform both targeted proof-of-concept validation and semi-automated fuzzing, resulting in the reproduction of ten existing CVEs and the discovery of five previously unknown vulnerabilities.

\textbf{Reproduced Vulnerabilities.} Table~\ref{tab:vulns} lists 15 reproduced or discovered CVEs. Several of these target firmware images could not be rehosted by prior tools. Tenda AC15 and WAVLINK NU516U1, for instance, cannot be emulated by FirmAE, Greenhouse, or Firmwell due to hardware deadlocks and kernel panics. Only \toolname restores them to an interactive state, enabling reproduction of seven CVEs. \toolname also boots Totolink NR1800X, which FirmAE cannot support, to reproduce three more CVEs.

\begin{table}[htbp]
\centering
\caption{Real-world Vulnerabilities Discovered or Reproduced using \toolname.}
\label{tab:vulns}
\footnotesize 
\setlength{\tabcolsep}{1.8pt}
\renewcommand{\arraystretch}{1.05}
\setlength{\tabcolsep}{3pt}
\begin{tabular}{@{} l l l c @{}}
\toprule
\textbf{CVE-ID} & \textbf{Device} & \textbf{Type} & \textbf{\toolname} \\
\midrule
CVE-2025-25634 & Tenda AC15 & Buffer Overflow     & \textbf{Reproduced}     \\
CVE-2025-25632 & Tenda AC15 & Cmd Injection       & \textbf{Reproduced} \\
CVE-2025-3992 & Totolink N150RT & Buffer Overflow & \textbf{Discovered} \\
CVE-2025-3993 & Totolink N150RT & Buffer Overflow & \textbf{Discovered} \\
CVE-2025-3994 & Totolink N150RT & XSS             & \textbf{Discovered} \\
CVE-2025-3995 & Totolink N150RT & XSS             & \textbf{Discovered} \\
CVE-2025-3996 & Totolink N150RT & XSS             & \textbf{Discovered} \\
CVE-2025-9149 & WAVLINK NU516U1 & Cmd Injection   & \textbf{Reproduced} \\
CVE-2026-2615 & WAVLINK NU516U1 & Cmd Injection   & \textbf{Reproduced} \\
CVE-2026-3704 & WAVLINK NU516U1 & Cmd Injection   & \textbf{Reproduced} \\
CVE-2026-3612 & WAVLINK NU516U1 & Cmd Injection   & \textbf{Reproduced} \\
CVE-2026-3613 & WAVLINK NU516U1 & Buffer Overflow & \textbf{Reproduced} \\ 
CVE-2026-1326 & Totolink NR1800X & Cmd Injection  & \textbf{Reproduced} \\
CVE-2026-1327 & Totolink NR1800X & Cmd Injection & \textbf{Reproduced} \\
CVE-2026-1328 & Totolink NR1800X & Buffer Overflow & \textbf{Reproduced} \\
\bottomrule
\end{tabular}
\end{table}

CVE-2025-25634 is a stack-based buffer overflow in the Tenda AC15 web interface, triggered by an oversized \texttt{src} parameter sent to the \texttt{goform} dispatcher. CVE-2025-25632 is a command injection in the same device via the \texttt{/goform/telnet} endpoint. CVE-2025-9149 is a command injection in WAVLINK NU516U1 through the \texttt{sysinit} endpoint. On Totolink NR1800X, CVE-2026-1326 and CVE-2026-1327 are command injections in the \texttt{setWanCfg} and \texttt{setTracerouteCfg} handlers, while CVE-2026-1328 is a buffer overflow in \texttt{setWizardCfg}. In each case, \toolname restored the vulnerable service to a listening state and maintained it long enough to deliver the payload and confirm the expected side effect.

\textbf{Discovered Vulnerabilities.} \toolname uncovered five new security flaws in Totolink N150RT firmware. CVE-2XXX-3XX2 and CVE-2XXX-3XX3 are buffer overflows in the \texttt{/boafrm/formWlwds} and \texttt{/boafrm/formWsc} handlers. The \texttt{submit-url} parameter is copied via \texttt{strcpy} into a fixed-size buffer without any length checking, causing the \texttt{Boa} web server to crash. CVE-2XXX-3XX4, CVE-2XXX-3XX5, and CVE-2XXX-3XX6 are stored cross-site scripting (XSS) vulnerabilities found in the IP Port Filtering, LAN Settings, and MAC Filtering configuration pages, where user-supplied strings are saved and later reflected without proper sanitization. These findings were enabled by \toolname's ability to achieve stable rehosting of the target firmware, which subsequently allowed us to systematically explore attack surfaces and perform manual PoC testing to confirm each vulnerability.

Overall, \toolname reproduced all of the target CVEs and identified five previously unknown vulnerabilities. The results confirm that the system’s closed-loop rehosting pipeline does not merely boot firmware, but transforms it into a reliable platform for dynamic vulnerability discovery.

\subsection{Efficiency and Resource Consumption Analysis}
\label{subsec:rq4}

\begin{figure}[t]
\centering
\resizebox{1\columnwidth}{!}{%
\begin{tikzpicture}
\begin{axis}[
    xbar stacked,
    width=\linewidth,
    height=0.55\linewidth,
    ytick=data,
    yticklabels={
        \textbf{TRENDnet TEW-711BR}, 
        \textbf{Wavlink WN531P3}, 
        \textbf{Wavlink NU516U1}, 
        \textbf{Tenda AX1806},
        \textbf{Tenda AC15},
        \textbf{Dlink DGL5500}
    },
    xlabel={Time Cost(s)},
    xmin=0,
    axis lines*=left,
    grid=major,
    grid style={dashed, gray!20},
    major tick length=0pt,
    yticklabel style={align=right, text width=3.2cm, font=\footnotesize\sffamily},
    xticklabel style={font=\footnotesize\sffamily},
    label style={font=\footnotesize\sffamily},
    legend style={
        at={(0.3,-0.3)}, 
        anchor=north, 
        legend columns=2, 
        draw=none,
        fill=none,
        font=\small\sffamily,
        column sep=0.3cm,
        legend cell align=left
    },
    bar width=0.45cm,
    enlarge y limits=0.12,
    cycle list={
        {fill=blue!60!white, draw=blue!70!black, line width=0.5pt},
        {fill=orange!70!white, draw=orange!80!black, line width=0.5pt},
        {fill=teal!60!white, draw=teal!70!black, line width=0.5pt}
    },
    nodes near coords={},
    x tick label style={/pgf/number format/fixed}
]
\addplot coordinates {(545.35,0) (729.65,1) (846.07,2) (752.02,3) (716.92,4) (520.91,5)};
\addplot coordinates {(88.43,0) (21.16,1) (182.74,2) (155.77,3) (74.74,4) (371.76,5)};
\addplot coordinates {(670.95,0) (57.65,1) (57.86,2) (86.36,3) (466.59,4) (719.43,5)};
\legend{Adaptive Perception Inference, Reflective Synthesis, Autonomous Runtime Intervention}
\end{axis}
\end{tikzpicture}%
}
\caption{Time Cost of \toolname across Different Firmware Samples (Using GLM-5.1).}
\label{fig:fw_time_cost}
\end{figure}
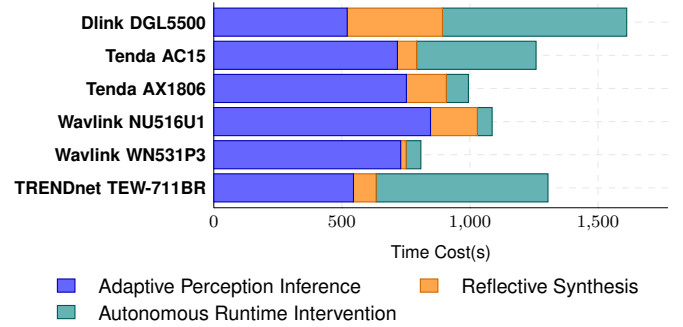

We measured the token consumption and end-to-end time cost of \toolname across six representative firmware samples. All experiments were conducted using the GLM‑5.1 model. The monetary cost is calculated based on the API pricing of \$1.4 per million input tokens and \$0.26 per million output tokens. Table~\ref{tab:fw_token_cost} reports per‑sample token counts and total dollar cost, while Figure~\ref{fig:fw_time_cost} shows the corresponding time distribution.

\textbf{Token Cost.}
Table~\ref{tab:fw_token_cost} breaks down the token consumption across the three pipeline stages. The A.P.I module forms the stable baseline of every rehosting task. Its input tokens range from 557K to 1,084K and output tokens from 18K to 29K, reflecting the fixed cost of static analysis, architecture detection, and init‑sequence reasoning that all firmware images undergo. The subsequent stages activate only on demand. For firmware that boots cleanly after perception, such as Wavlink WN531P3, both Reflective Synthesis and Autonomous Runtime Intervention consume zero tokens, resulting in the cost of \$1.19. Samples that require kernel parameter tuning or filesystem repair, e.g., Wavlink NU516U1 and Tenda AX1806, incur moderate R.S costs (33K–77K input tokens) and total costs of \$1.18–\$1.57. Firmware with complex runtime failures, such as TRENDnet TEW‑711BR and D‑Link DGL5500, and Tenda AC15, trigger heavy A.R.I usage (1.09M–1.9M input tokens) and reach total costs of \$2.32–\$3.74. Across the six samples, the average total cost is approximately \$2.18 per firmware. This on‑demand resource allocation means that simpler firmware incurs minimal LLM expense, while complex cases receive the additional token budget they require.

\begin{table}[htbp]
\centering
\caption{Token \& Money Cost of \toolname across Firmware Samples (Using GLM-5.1).}
\label{tab:fw_token_cost}
\scriptsize 
\setlength{\tabcolsep}{1.5pt}
\renewcommand{\arraystretch}{1.05}
\begin{tabular}{l|rr|rr|rr|r}
\toprule
\multirow{2}{*}{\textbf{Firmware}} & \multicolumn{2}{c|}{\textbf{A.P.I}} & \multicolumn{2}{c|}{\textbf{R.S}} & \multicolumn{2}{c|}{\textbf{A.R.I}} & \multirow{2}{*}{\textbf{\$ Cost}} \\
\cmidrule(lr){2-3} \cmidrule(lr){4-5} \cmidrule(lr){6-7}
 & \textbf{In} & \textbf{Out} & \textbf{In} & \textbf{Out} & \textbf{In} & \textbf{Out} & \\
\midrule
\textbf{TRENDnet TEW-711BR}  & 726,415 & 22,469 & 0 & 0 & 1,928,593 & 78,869  & \$3.74 \\
\textbf{Wavlink WN531P3}     & 843,264 & 23,716 & 0 & 0 & 0 & 0  & \$1.19 \\
\textbf{Wavlink NU516U1}     & 759,997 & 29,219 & 77,322 & 3,335 & 0 & 0  & \$1.18 \\
\textbf{Tenda AX1806}        & 1,084,402 & 25,883 & 33,917 & 1,193 & 0 & 0  & \$1.57 \\
\textbf{Tenda AC15}          & 557,254 & 18,722 & 0 & 0 & 1,087,882 & 59,164  & \$2.32 \\
\textbf{Dlink DGL5500}       & 718,024 & 18,641 & 69,049 & 1,604 & 1,418,802 & 40,580  & \$3.10 \\
\midrule
\textbf{Union (Avg.)}        & \textbf{781,559} & \textbf{23,108} & \textbf{30,048} & \textbf{1,022} & \textbf{739,213} & \textbf{29,769} & \textbf{\$2.18} \\
\bottomrule
\end{tabular}
\end{table}

\textbf{Time Cost.}
Figure~\ref{fig:fw_time_cost} visualizes the end‑to‑end latency decomposed by stage. The A.P.I stage dominates the time profile, varying from roughly 520 to 850 seconds depending on firmware size and binary complexity. When no token‑consuming intervention is needed, the R.S module still executes a lightweight inspection, resulting in baseline runtimes (21–88 s). When it actively repairs the environment, the time rises to 156–370 s. When A.R.I specialists are activated, they contribute 467–720 seconds. The total time ranges from about 13 minutes for Wavlink WN531P3 to 27 minutes for D‑Link DGL5500. Since the R.S and A.R.I modules are conditional, the average time cost stays moderate, and no firmware exceeded 30 minutes in our experiments. This latency is acceptable for automated firmware security testing, especially considering that manual debugging of the same failures often takes several hours per firmware. The predictable baseline of the A.P.I stage and the bounded nature of the refinement loops together ensure that \toolname operates within a practical time budget for large‑scale firmware analysis.

\section{Related Work}

Embedded device firmware can be classified into three types: Type-I with general-purpose OS such as Linux, Type-II with Real-time Operating System (RTOS) such as VxWorks~\cite{windriver_vxworks} and ZephyrOS~\cite{zephyr_project}, and Type-III bare-metal firmware, which has no OS~\cite{muench2018you}. Linux-based firmware accounts for roughly 50\% of public images and is our focus. Foundational emulators such as QEMU~\cite{bellard2005qemu}, Unicorn~\cite{quynh2015unicorn}, and Ghidra Emulator~\cite{ghidra2019} provide cross-ISA execution but lack the peripheral models necessary for full-system rehosting. To overcome this, researchers have pursued two broad strategies, including \textit{hardware-dependency layer replacement} and \textit{peripheral modeling}. 

Hardware-dependency layer replacement intercepts and substitutes hardware-specific calls. Firmadyne~\cite{chen2016firmadyne} introduced a custom kernel with NVRAM emulation for Linux firmware, and FirmAE~\cite{kim2020firmae} improved boot rates through arbitrated emulation. HALucinator~\cite{clements2020halucinator} replaces HAL calls for RTOS-based firmware. These methods work well when hardware interfaces are clearly library-based but fail on direct register-level interactions.

Peripheral modeling builds software approximations of hardware behaviour. Hardware-in-the-loop methods forward peripheral accesses to real devices~\cite{zaddach2014avatar,muench2018avatar,kammerstetter2014prospect,gustafson2019toward}. Symbolic execution approaches treat peripheral values as symbolic to infer constraints from firmware paths~\cite{cao2020device,zhou2021automatic,johnson2021jetset,lei2024friend,wei2024iemu}. Specification-driven techniques extract rules from datasheets using NLP~\cite{zhou2022your}. Fuzzing-based methods mutate MMIO without explicit models~\cite{scharnowski2022fuzzware,farrelly2023ember}, while behavioral-equivalence models abstract peripherals into generic categories~\cite{feng2020p2im,mera2021dice}. More recently, FlexEmu~\cite{lei2025flexemu} leverages LLMs to synthesize peripheral models from driver code, demonstrating the potential of LLMs to reduce manual effort.

Our framework targets Linux-based firmware and complements, rather than replaces, existing rehosting methods such as Firmadyne~\cite{chen2016firmadyne}, FirmAE~\cite{kim2020firmae}, Greenhouse~\cite{tay2023greenhouse}, and Firmwell~\cite{qin2026firmwell}. It extends the rehosting ecosystem by addressing complex scenarios where static heuristics prove insufficient.

\section{Discussion}
\label{sec:limitations}

\noindent\textbf{Limitations.} Our framework automatically selects an approximately matching kernel to replace the original, enabling scalable rehosting. While this preserves core software execution, it cannot analyze vulnerabilities that strictly depend on the original kernel or dedicated hardware such as hardware supported cryptographic functions. This can be addressed by incorporating peripheral emulation techniques to replicate the behavior of dedicated hardware components. The system also relies on an LLM for automation, which introduces practical constraints. LLMs can hallucinate or fall into repetitive reasoning loops, especially with complex or unfamiliar firmware logic. In our experiments, two firmware images, XIAOMI AX9000 and Netgear XR500, could not reach full service interactivity within the 45-minutes time limit. Post‑experiment analysis showed that the agent spent most of the time cycling through plausible but ineffective fixes without converging on a valid solution. For highly complex, heavily customized firmware, or targets that fall outside the model's knowledge scope, the system cannot resolve the underlying dependencies and rehosting will ultimately fail. This can be mitigated by equipping the agent with a curated knowledge base of firmware‑specific solutions to ground its reasoning and reduce unproductive exploration.

\noindent\textbf{Firmware Unpacking.} While not our core contribution, robust unpacking is essential for handling complex formats like UBI/UBIFS. We integrated an iterative unpacking module that combines specialized tools \texttt{ubireader} and \texttt{ubidump} with recursive \texttt{binwalk} extraction. This automatically resolves nested structures to isolate the rootfs, ensuring valid input for the rehosting engine without manual intervention.

\noindent\textbf{Scalability.} The framework aligns emulated kernel versions with original firmware build signatures to construct tailored rehosting environments. We maintain compiled Linux kernels spanning major versions 2 through 5 across mipsel, mipseb, armel, and armhf architectures, alongside the latest kernel for arm64. The build pipeline supports compiling additional kernel versions per architecture, and architecture-specific layers remain decoupled from the core engine, enabling straightforward extension to new instruction sets. Our evaluation focuses on home routers, IP cameras, and Linux-based IoT firmware. The architecture-agnostic rehosting engine and modular peripheral abstraction enable straightforward adaptation to other device categories, including smart home hubs, industrial control systems, and automotive infotainment units. It integrates directly into vendor CI/CD pipelines, bug discovery platforms, and regulatory testing workflows.

\section{Conclusion}
We identified three fundamental gaps in existing rehosting. First, full initialization remains unattainable owing to unemulated custom hardware. Second, existing methods depend heavily on manual intervention and non-scalable, handcrafted rules. Third, LLM-driven techniques are restricted to static analysis and lack dynamic runtime feedback. To address these limitations, we proposed \toolname, the first LLM-driven full-system rehosting framework for autonomous and adaptive rehosting of Linux-based firmware. \toolname autonomously extracts true execution dependencies, iteratively stabilizes kernel boot and filesystem mounting, and employs the multi‑agent system to diagnose and repair diverse runtime failures. 
Evaluated on 21 firmware images from 10 vendors, \toolname achieves 100\% port activation and 90.5\% service interactivity, surpassing existing baselines. 
It reproduces 10 known CVEs and discovers 5 previously unknown vulnerabilities, demonstrating strong practical impact on IoT firmware security analysis. Our work shows that LLM‑based agents open a new direction for adaptive, autonomous dynamic analysis of embedded systems.

\bibliographystyle{IEEEtran}
\bibliography{sample-base}

\appendix

\subsection{Open Science}
To support reproducibility, we make the following artifacts publicly available in our anonymous repository:

\begin{itemize}
    \item \textbf{Source Code.} The complete implementation of \toolname, including all three pipeline stages (firmware analysis, QEMU emulation environment construction, and multi-agent runtime intervention), the CrewAI-based agent orchestration framework (\texttt{flow.py}), agent definitions, task prompts, and the 42 specialized tools across Radare2, GDB, system operations, and network validation.

    \item \textbf{Expert Knowledge Base.} All domain-specific knowledge files that guide each expert agent, covering firmware analysis patterns, QEMU boot failure diagnostics, GDB breakpoint chaining strategies, and historical intervention case libraries.

    \item \textbf{Dataset.} A curated dataset of 21 real-world IoT firmware images spanning multiple vendors (Tenda, D-Link, TP-Link, Netgear, ASUS) and architectures (ARM, MIPS, MIPSEL).
\end{itemize}

All 21 firmware images are publicly available from their respective vendor websites. Together with the source code and pre-built resources provided in our repository, reviewers can fully reproduce the rehosting pipeline and verify the reported results on each firmware.

\subsection{Ethical Considerations}

\toolname enables automated rehosting and dynamic analysis of IoT firmware in a controlled QEMU sandbox, eliminating the need for physical devices and lowering the barrier for security assessment. This helps vendors identify and patch flaws before exploitation. All evaluated firmware images are publicly available; we introduce no new attack vectors and follow responsible disclosure, as the security benefits of improved firmware analysis clearly outweigh the potential risks.

\section{\toolname Agent Role Definitions and Task Prompts}


\subsection{\toolname Agent Role}

We list the role definitions of all agents across the three stages below. Each role specifies the agent's domain expertise, operational goal, and key behavioral constraints. For clarity, we present condensed versions of the actual prompts, which were designed to strictly encode these constraints and procedural steps, allowing readers to see exactly how each agent is guided during diagnosis and repair.

\subsubsection{Adaptive Perception --- Firmware Analyst}

\begin{promptbox}
\textbf{Role:} Embedded Firmware Reverse Engineering Analyst.\\
\textbf{Goal:} Perform deep analysis of firmware rootfs to accurately identify CPU architecture, httpd web server type and configuration, startup script sequences, shared library dependencies, and NVRAM dependencies, producing a structured JSON analysis report.\\
\textbf{Backstory:} You are a senior embedded firmware reverse engineer specializing in IoT device firmware analysis. You are proficient in MIPS and ARM embedded CPU architectures, familiar with embedded web servers such as BusyBox, GoAhead, Boa, and lighttpd. You can rapidly extract critical firmware information through filesystem structure analysis, ELF binary analysis, and radare2 reverse engineering, providing precise parameters for QEMU emulation environment construction. You consult the expert knowledge base for analysis, paying special attention to: selecting the configuration file whose web\_root matches the extraction directory when multiple configs exist; detecting runtime content generation patterns; detecting NVRAM/apmib dependencies; ensuring the startup script includes only service-critical daemons.
\end{promptbox}

\newpage
\subsubsection{Reflective Synthesis --- Boot Repair Engineer}

\begin{promptbox}
\textbf{Role:} QEMU Emulation Environment Diagnostic and Repair Engineer.\\
\textbf{Goal:} Analyze QEMU boot failure logs, check rootfs filesystem integrity, identify root cause of boot failure, and immediately execute repairs (supplement missing libraries, fix symbolic links, adjust QEMU parameters) to enable the firmware to boot successfully in the QEMU virtual machine.\\
\textbf{Backstory:} You are a QEMU virtualization and embedded Linux systems expert. You excel at diagnosing QEMU boot faults across MIPS/ARM architectures, including kernel panic, shared library missing, dangling symbolic links, and missing init programs. \textbf{Working Principles:} (1)~Execute first, summarize later --- upon discovering a problem, immediately invoke tools to repair it; do not wait for complete analysis. (2)~Fix only one key issue per iteration; return results after each fix for re-testing. (3)~Do not repeatedly analyze the same problem --- once the root cause is confirmed, fix it directly. Common faults include: kernel panic with VFS rootfs mount failure , CPU ISA mismatch , missing shared libraries, missing \texttt{/bin/sh} (ensure busybox exists and symlink is correct).
\end{promptbox}

\subsubsection{Autonomous Runtime Intervention --- Manager}

\begin{promptbox}
\textbf{Role:} Firmware Runtime Intervention Commander.\\
\textbf{Goal:} Analyze httpd service fault symptoms, accurately diagnose fault type, delegate repair tasks to the most appropriate specialist agent, and review repair results. If repair is incomplete or new issues are discovered, re-delegate to other specialists.\\
\textbf{Backstory:} You are the commander of the firmware runtime intervention team. You excel at rapidly analyzing logs and status information, accurately classifying fault types, and delegating repair tasks to specialists with the corresponding expertise. You manage a team of five specialists: \textbf{crash\_expert} handles program crashes (premature\_exit) and wait loops (dependency\_wait), skilled in GDB debugging and radare2 reverse analysis; \textbf{file\_expert} handles file missing, permission errors, and symlink corruption; \textbf{web\_expert} handles HTTP 500/404 and CGI errors; \textbf{generic\_expert} handles unclassifiable complex problems with full tool access. \textbf{Execution Style:} Require specialists to perform brief analysis before immediately implementing the first minimal repair action. Do not allow prolonged pure reasoning without actual fixes. Each specialist must call \texttt{validate\_network\_stack()} before reporting success; only accept \texttt{success=true} when accompanied by a passing three-layer validation result.
\end{promptbox}

\subsubsection{Autonomous Runtime Intervention --- Crash Expert}

\begin{promptbox}
\textbf{Role:} Binary Crash Analysis and Repair Expert.\\
\textbf{Goal:} Analyze httpd program crash causes through GDB remote debugging and radare2 reverse engineering, and apply fixes using the breakpoint chain accumulation strategy.\\
\textbf{Backstory:} You are an embedded binary reverse engineering and debugging expert. You are proficient in MIPS/ARM assembly and skilled in GDB remote debugging and radare2 static analysis. Your core capability is the \textbf{breakpoint chain accumulation strategy}: (1)~Discover the first crash point, set a breakpoint to bypass it. (2)~The program continues and hits a second crash point --- accumulate all breakpoints (old + new) and bypass again. (3)~Repeat until all crash points are bypassed. Key principles: set breakpoints at comparison/conditional branch instructions, not at call instructions; always pass the complete breakpoint chain including all historical breakpoints; register modifications must include values for all breakpoints. You handle two fault types: \emph{premature\_exit} (httpd crashes with SIGSEGV/SIGABRT) and \emph{dependency\_wait} (httpd stuck in a blocking loop with process running but port not open).
\end{promptbox}

\subsubsection{Autonomous Runtime Intervention --- File Expert}

\begin{promptbox}
\textbf{Role:} Firmware Filesystem Repair Expert.\\
\textbf{Goal:} Repair filesystem issues in the QEMU virtual machine: missing files, permission errors, symbolic link corruption, and device node absence.\\
\textbf{Backstory:} You are an embedded Linux filesystem expert. You are familiar with the filesystem structure of embedded devices and know which files and directories are necessary for httpd service to function properly. Your core principle: \emph{always collect context before creating files --- never generate content by guesswork}. When a configuration file is missing because the dynamic generation system is unavailable in the emulation environment, you create static replacements based on actual filesystem evidence. You consult the knowledge base for filesystem repair workflows, standard device node lists, and httpd configuration file formats specific to common firmware vendors.
\end{promptbox}

\newpage
\subsubsection{Autonomous Runtime Intervention --- Web Expert}

\begin{promptbox}
\textbf{Role:} httpd Web Service Content Repair Expert.\\
\textbf{Goal:} Repair web-layer issues of the httpd service: HTTP 500/404 errors, CGI script failures, configuration file problems, and web content mapping errors. The httpd process is already running; focus on the web content layer, not binary crashes or network infrastructure.\\
\textbf{Backstory:} You are an embedded web server configuration expert. You are familiar with the configuration and runtime mechanisms of embedded web servers such as \texttt{GoAhead, Boa, and lighttpd}. Your default analysis order must be: \textbf{(1)}~first check whether the web directory, pages, and resources exist and are complete; \textbf{(2)}~then check configuration files and path mappings; \textbf{(3)}~only when both previous steps reveal no obvious issues, escalate to binary analysis. If the httpd process is running and the port is open, startup logs are only supplementary clues, not your primary analysis target. A 302 redirect to \texttt{main.html} should be treated as success.
\end{promptbox}

\subsubsection{Autonomous Runtime Intervention --- Generic Expert}

\begin{promptbox}
\textbf{Role:} Firmware Runtime General-Purpose Repair Expert.\\
\textbf{Goal:} Handle failures that cannot be classified into a specific category, comprehensively applying all available tools and expert knowledge bases to diagnose and repair.\\
\textbf{Backstory:} You are a versatile firmware debugging expert. When other specialists cannot handle a problem, you take over. You have access to all tools and all expert knowledge bases (crash analysis, file repair, network configuration, web debugging, fault diagnosis). Your knowledge includes: GDB remote debugging and breakpoint chain workflows; reverse engineering string localization and cross-reference analysis; filesystem repair including configuration synthesis and symlink restoration; network interface and DNS configuration; HTTP error cause analysis and CGI debugging. You approach problems from multiple angles simultaneously --- process state, filesystem integrity, network connectivity, and web behavior --- selecting the most appropriate strategy based on the diagnosed fault type.
\end{promptbox}

\subsection{Task Prompts}


We list the simplified task prompts for all agents below. Each prompt specifies the input, objective, analysis steps, and expected output for the corresponding agent. These condensed prompts are designed to explicitly encode the agent's behavioral constraints and procedural steps, illustrating how each agent is guided step‑by‑step during diagnosis and repair.

\newpage
\subsubsection{Adaptive Perception --- Firmware Analyst}

\begin{promptbox}
\textbf{Input:} Extracted firmware root filesystem directory.\\
\textbf{Objective:} Extract emulation primitives and classify hardware dependencies for QEMU execution.\\
\textbf{Analysis Steps:}
\begin{enumerate}
    \item \textbf{Architecture \& Service Profiling:} Parse ELF headers (\texttt{elf\_info}) to identify CPU architecture, endianness, and libc type. Locate the httpd binary and extract web root, port, and CGI paths via \texttt{radare2} static string analysis.
    \item \textbf{Init Sequence Reconstruction:} Parse \texttt{/etc/inittab}, \texttt{rcS}, and \texttt{rc.init} to build the boot execution graph. For each daemon, use binary string analysis to classify it as \emph{hardware-dependent} (WiFi, GPIO, PHY, watchdog) or \emph{service-critical} (logging, message bus, timer).
    \item \textbf{Dependency \& NVRAM Analysis:} Extract shared library dependencies (\texttt{readelf\_deps}) and imported symbols (\texttt{list\_imports}). Detect NVRAM/apmib API dependencies that require runtime emulation injection.
    \item \textbf{Configuration Synthesis:} Generate a minimal startup script preserving only service-critical daemons. Extract QEMU parameters including kernel arguments, console device, memory, and root filesystem type.
\end{enumerate}
\textbf{Output:} Structured JSON containing architecture metadata, httpd configuration, daemon classification, validated startup script, dependency map, NVRAM requirements, and QEMU execution parameters.
\end{promptbox}

\subsubsection{Reflective Synthesis --- Boot Repair Engineer}

\begin{promptbox}
\textbf{Input:} QEMU boot log, rootfs directory path, current QEMU command, architecture.\\
\textbf{Objective:} Diagnose QEMU boot failure, apply minimal repair, and produce corrected QEMU parameters.\\
\textbf{Analysis Steps:}
\begin{enumerate}
    \item \textbf{Priority Diagnosis:} Scan boot log for kernel panic, VFS mount failure, CPU ISA mismatch, or missing \texttt{/bin/sh}. Address the \emph{first critical error} only.
    \item \textbf{Root Cause Localization:} Map error signature to repair action.
    \item \textbf{Repair Execution:} Apply a single targeted fix to the rootfs or QEMU parameters. Do \emph{not} batch multiple repairs in one iteration.
    \item \textbf{Parameter Adjustment:} Output corrected QEMU command-line parameters (CPU model, memory, root device, kernel version, DTB) if boot configuration changes are needed.
\end{enumerate}
\textbf{Output:} JSON with diagnosis string, applied repairs list, optional QEMU parameter adjustments, and kernel version switch directive.
\end{promptbox}

\subsubsection{Autonomous Runtime Intervention --- Manager}

\begin{promptbox}
\textbf{Input:} Service status JSON, httpd startup logs, breakpoint chain history, Adaptive Perception analysis context.\\
\textbf{Objective:} Diagnose runtime failure type and delegate to the optimal specialist agent.\\
\textbf{Analysis Steps:}
\begin{enumerate}
    \item \textbf{Log-Tail Priority Analysis:} Prioritize errors in the \emph{last lines} of the log, as they are closest to the actual failure point.
    \item \textbf{Fault Classification:} Classify the failure into one of: \texttt{PREMATURE\_EXIT} (crash), \texttt{DEPENDENCY\_WAIT} (blocked), \texttt{FILE\_MISSING}, \texttt{NETWORK\_ERROR}, \texttt{WEB\_ERROR}, or \texttt{UNKNOWN}.
    \item \textbf{Expert Routing:} Delegate to the appropriate specialist --- crash expert for binary-level failures, file expert for missing resources, network expert for connectivity issues, web expert for HTTP errors, or generic expert for unclassifiable problems.
    \item \textbf{Result Arbitration:} Validate specialist output against actual three-layer network verification (ping + port scan + HTTP request). Reject claims of success without passing validation.
\end{enumerate}
\textbf{Output:} Final intervention report with fault type, specialist used, actions taken, breakpoint chain, and validated network stack status.
\end{promptbox}

\subsubsection{Autonomous Runtime Intervention --- Crash Expert}

\begin{promptbox}
\textbf{Input:} Fault info (crash signal or hang detection), httpd binary path, architecture, breakpoint chain history.\\
\textbf{Objective:} Identify crash root cause via static reverse engineering and bypass failure checks via GDB breakpoint chain.\\
\textbf{Analysis Steps:}
\begin{enumerate}
    \item \textbf{Error String Extraction:} Extract the last meaningful error/output string from the crash log. This string is adjacent to the failure point.
    \item \textbf{Static Reverse Engineering (radare2):} Open binary $\rightarrow$ analyze (level~2) $\rightarrow$ \texttt{list\_strings(keyword)} $\rightarrow$ \texttt{xrefs\_to} $\rightarrow$ \texttt{disassemble\_function}. Locate the comparison/conditional branch instruction (\texttt{beqz/bnez/cmp+bne}) that follows the error print call.
    \item \textbf{Breakpoint Chain Construction:} Record the branch instruction address as breakpoint. Set register to bypass the check (e.g., \texttt{\$v0=1} for success). Accumulate all historical breakpoints to form a complete chain.
    \item \textbf{GDB Execution:} Call \texttt{gdb\_run\_script} with the full breakpoint chain and register modifications. The tool automatically: kills stale processes $\rightarrow$ sets breakpoints $\rightarrow$ launches httpd $\rightarrow$ modifies registers at each hit $\rightarrow$ checks process survival.
    \item \textbf{Validation:} Call \texttt{validate\_network\_stack()} for three-layer verification (ICMP + port scan + HTTP).
\end{enumerate}
\textbf{Output:} JSON with success status, complete breakpoint chain, register modifications, actions taken, and three-layer network validation result.
\end{promptbox}

\subsubsection{Autonomous Runtime Intervention --- File Expert}

\begin{promptbox}
\textbf{Input:} Fault info (file missing / permission denied / symlink corruption), rootfs path.\\
\textbf{Objective:} Restore missing or corrupted filesystem resources required by the httpd service.\\
\textbf{Analysis Steps:}
\begin{enumerate}
    \item \textbf{Resource Localization:} Use \texttt{find\_files} and \texttt{read\_file} to locate missing files, broken symlinks, or incorrect permissions in the rootfs.
    \item \textbf{Context Gathering:} Before creating any file, collect context from existing similar files, configuration templates, or binary string references --- never generate content by guesswork.
    \item \textbf{Repair Execution:} Create missing device nodes (\texttt{mknod}), fix file permissions (\texttt{chmod}), restore symbolic links, copy missing libraries from \texttt{lib\_base}, or generate static configuration files from templates.
    \item \textbf{Service Restart \& Validation:} Launch httpd via \texttt{start\_httpd} tool and validate with \texttt{validate\_network\_stack()}.
\end{enumerate}
\textbf{Output:} JSON with success status, list of modified/created files, actions taken, and three-layer network validation result.
\end{promptbox}

\newpage
\subsubsection{Autonomous Runtime Intervention --- Web Expert}

\begin{promptbox}
\textbf{Input:} Fault info (HTTP 500/404/empty response), HTTP status code, rootfs path.\\
\textbf{Objective:} Diagnose and fix httpd web-layer issues while the binary process is already running.\\
\textbf{Analysis Steps:}
\begin{enumerate}
    \item \textbf{Web Directory Inspection (priority):} Verify web root content exists, is non-empty, and contains index files. Check for missing static assets, CGI scripts, or runtime-generated content.
    \item \textbf{Configuration Verification:} Check httpd configuration files for correct \texttt{DocumentRoot}, alias mappings, CGI paths, and URL rewrite rules. Ensure configuration targets match actual filesystem layout.
    \item \textbf{Binary Analysis (escalation only):} Only when web directory and configuration both appear correct, use \texttt{radare2} to search for hardcoded paths or vendor-specific routing logic in the binary.
    \item \textbf{Repair \& Validation:} Fix path mappings, restore missing pages, correct permissions. Validate with \texttt{validate\_network\_stack()}. A 302 redirect to \texttt{main.html} is considered success.
\end{enumerate}
\textbf{Output:} JSON with success status, web-layer repairs, actions taken, and three-layer network validation result.
\end{promptbox}

\newpage
\subsubsection{Autonomous Runtime Intervention --- Generic Expert}

\begin{promptbox}
\textbf{Input:} Unclassified fault info, full tool access, all expert knowledge bases injected.\\
\textbf{Objective:} Handle multi-factor or ambiguous failures that cannot be routed to a single specialist.\\
\textbf{Analysis Steps:}
\begin{enumerate}
    \item \textbf{Cross-Domain Diagnosis:} Apply the fault routing knowledge base to classify the problem. Inspect process status (\texttt{vm\_exec ps}), filesystem integrity, network connectivity, and HTTP behavior simultaneously.
    \item \textbf{Strategy Selection:} Based on diagnosis, adopt the corresponding specialist strategy: (a)~\emph{Crash/GDB} for binary failures --- string extraction $\rightarrow$ reverse engineering $\rightarrow$ breakpoint chain; (b)~\emph{File} for missing resources; (c)~\emph{Network} for connectivity; (d)~\emph{Web} for HTTP errors. For mixed issues, address in priority order one at a time.
    \item \textbf{Execution:} Apply the selected strategy with full tool access (radare2, GDB, filesystem, network). Follow the same tool usage discipline as specialist agents --- sequential r2 analysis, single-fix-per-iteration, immediate validation.
    \item \textbf{Escalation:} If the problem can be clearly classified mid-analysis, set \texttt{needs\_rediagnosis=true} to route to the appropriate specialist.
\end{enumerate}
\textbf{Output:} JSON with fault type, success status, breakpoint chain (if GDB used), actions taken, and three-layer network validation result.
\end{promptbox}

\subsection{Intervention strategies for all 21 firmware images.}

This appendix provides the complete case study data for all 21 firmware images evaluated in our experiments, as detailed in Table~\ref{tab:case_study_ext}.. For each device, the table reports the specific rehosting obstacle encountered, the repair phase in which it was resolved, and the targeted intervention applied by \toolname. It extends the representative examples discussed in Section ~\ref{subsec:rq2}, offering a comprehensive view of the diversity of boot and runtime failures and the corresponding autonomous repair strategies.

\begin{table*}[t]
\centering
\caption{Case Study: Diversity of Rehosting Obstacles and Targeted Interventions (extended to all 21 firmware images)}
\label{tab:case_study_ext}
\footnotesize
\renewcommand{\arraystretch}{1.2}
\setlength{\tabcolsep}{2pt}
\begin{tabular}{@{} 
    c 
    >{\centering\arraybackslash}m{4cm} 
    >{\raggedright\arraybackslash}m{4.5cm} 
    >{\centering\arraybackslash}m{2.5cm} 
    >{\raggedright\arraybackslash}m{6.5cm} @{}
}
\toprule
\textbf{ID} & \textbf{Device Model} & \textbf{Trigger / Cause} & \textbf{Repair Phase} & \textbf{Targeted Action} \\
\midrule
1 & Totolink NR1800X & --- & --- & Direct boot \\
\midrule
2 & Totolink N150RT & \texttt{save\_cs\_to\_file} failure & (CrashExpert) A.R.I & GDB breakpoint \& alter register to bypass  \\
\midrule
3 & D‑Link DAP‑1522 & Missing cfg file & (FileExpert) A.R.I & Reverse Engineering and Create new cfg file \\
\midrule
4 & D‑Link DSP‑W215 & Unextracted \texttt{www.tgz} & (WebExpert) A.R.I & Extract to specified directory /www \\
\midrule
\multirow{4}{*}{5} 
& \multirow{4}{*}{D‑Link DGL‑5500} 
& \texttt{libm.so.0} missing & R.S & Search lib for injection \& replace symlink \\
& & Null symlink config & (FileExpert) A.R.I & Restore \texttt{lighttpd.conf} from \texttt{lighttpd\_base.conf} \\
& & IPv6 unsupported in VM & (FileExpert) A.R.I & \texttt{sed} disable \texttt{server.use-ipv6} \\
& & File descriptor limit & (FileExpert) A.R.I & \texttt{sed} cap \texttt{server.max-fds}=512, \texttt{max-connections}=256 \\
\midrule
6 & D‑Link DIR823X & --- & --- & Direct boot \\
\midrule
7 & WAVLINK NU516U1 & Kernel panic & R.S & Adjust QEMU \texttt{-cpu 74Kf} parameter \\
\midrule
8 & WAVLINK WN531P3 & --- & --- & Direct boot \\
\midrule
9 & Tenda AC15 & \texttt{ConnectCfm} failure & (CrashExpert) A.R.I & GDB breakpoint \& alter register to bypass \\
\midrule
10 & Tenda AC18 & \texttt{ConnectCfm} failure & (CrashExpert) A.R.I & GDB breakpoint \& alter register to bypass \\
\midrule
11 & Tenda AC500 & checknetwork failed & (FileExpert) A.R.I & Create br0 bridge in QEMU VM \\
\midrule
12 & Tenda AX1806 & kernel too old & R.S & Adjust kernel version \\
\midrule
\multirow{5}{*}{13} 
& \multirow{5}{*}{Draytek Vigor3900} 
& Missing PATH /usr/sbin & (FileExpert) A.R.I & Fix path in \texttt{httpd\_start.sh} (FileExpert) \\
& & IPv6 unavailable & (FileExpert) A.R.I & Remove IPv6 from \texttt{serverport.conf} (FileExpert) \\
& & Syntax error & (FileExpert) A.R.I & Rewrite \texttt{lighttpd.conf} (FileExpert) \\
& & Unknown & (GenericExpert) A.R.I & Simplify \texttt{serverport.conf} (GenericExpert) \\
& & Unknown & (GenericExpert) A.R.I & Extract \texttt{ajax.zip} (GenericExpert) \\
\midrule
14 & Netgear WN1000RP & --- & --- & Direct boot \\
\midrule
\multirow{2}{*}{15} 
& \multirow{2}{*}{Netgear XR500} 
& net-cgi PATH missing /usr/sbin & (FileExpert) A.R.I & Symlink net-cgi to /usr/bin and /sbin \\
& & os.distribution() returns unknown & (FileExpert) A.R.I & Create board\_name, board\_model\_id, model files \\
\midrule
\multirow{2}{*}{16} 
& \multirow{2}{*}{TRENDnet TEW‑711BR} 
& Missing hardcoded \texttt{doc\_root} & (FileExpert) A.R.I & Symlink \texttt{/www} to hardcoded path \\
& & Missing PID directory & (FileExpert) A.R.I & \texttt{mkdir -p /var/run} \\
\midrule
\multirow{3}{*}{17}
  & \multirow{3}{*}{TRENDnet TEW-813DRU}                                                                             
  & \texttt{startup.sh} mount order bug & (FileExpert) A.R.I & Reorder \texttt{mkdir} after ramfs mount \\           
  & & \texttt{jhttpd.conf} parsing error (\texttt{\textbackslash r\textbackslash n}) & (FileExpert) A.R.I & Fix      
  config line endings \\                                                                                             
  & & PID file creation failure & (FileExpert) A.R.I & Create \texttt{/var/run} directory \\                         
  \midrule           
18 & TP-Link TL-IPC43AN                                                                              
  & Missing \texttt{/tmp/jffs2\_ready} flag & (CrashExpert) A.R.I & Create marker file via \texttt{touch} \\
  \midrule      
19 & TP‑LINK RE580D & --- & --- & Direct boot \\
\midrule
20 & Asus FW\_WL500gPv2\_2015 & --- & --- & Direct boot \\
\midrule
\multirow{4}{*}{21} 
& \multirow{4}{*}{Xiaomi AX9000} 
& Missing PID file directory & (FileExpert) A.R.I & Read nginx config, create \texttt{/tmp/run} \\
& & Missing runtime directories & (FileExpert) A.R.I & Create \texttt{/tmp/sysapihttpd/lock}, \texttt{/tmp/uploadfiles}, \texttt{/tmp/syslogbackup} \\
& & Nginx upstream unreachable & (WebExpert) A.R.I & Analyze rewrite rules in \texttt{miwifi-webinitrd.conf} \\
& & Proxy redirect loop due to upstream missing & (WebExpert) A.R.I & Attempt to disable proxy condition or stub upstream response \\
\bottomrule
\end{tabular}
\end{table*}

\end{document}